\documentclass{article}
\usepackage[utf8]{inputenc}
\usepackage[margin=1.1in]{geometry}
\usepackage{xcolor}
\usepackage{amsmath,amssymb,amsthm,float,tkz-berge,caption,mathtools}
\usepackage{cancel}
\usepackage{subcaption}
\usepackage{algorithm}
\usepackage[noend]{algpseudocode}

\newcommand{\E}{\mathbb{E}}
\newcommand{\p}{\mathbb{P}}
\newcommand{\OO}{\mathcal{O}}

\DeclareMathOperator*{\argmax}{arg\,max}

 \newtheorem{theorem}{Theorem}
 \newtheorem*{theorem*}{Theorem}
 \newtheorem{lemma}{Lemma}
 \newtheorem{definition}{Definition}
 \newtheorem*{definition*}{Definition}
 \newtheorem*{lemma*}{Lemma}
 \newtheorem{corollary}{Corollary}
 \newtheorem*{corollary*}{Corollary}
 
 \newtheorem*{claim*}{Claim}
 \newtheorem{proposition}{Proposition}
 \newtheorem*{proposition*}{Proposition}
\allowdisplaybreaks
\title{Learning Graphs from Noisy Epidemic Cascades}
\author{Jessica Hoffmann, Constantine Caramanis}
\date{\today}

\begin{document}
	\maketitle

\begin{abstract}
        We consider the problem of learning the weighted edges of a graph by observing the noisy times of infection for multiple epidemic cascades on this graph. Past work has considered this problem when the cascade information, i.e., infection times, are known exactly. Though the noisy setting is well motivated by many epidemic processes (e.g., most human epidemics), to the best of our knowledge, very little is known about when it is solvable. Previous work on the no-noise setting critically uses the ordering information. If noise can reverse this -- a node's reported (noisy) infection time comes after the reported infection time of some node it infected -- then we are unable to see how previous results can be extended.
        We therefore tackle two versions of the noisy setting: the {\em limited-noise} setting, where we know noisy times of infections, and the {\em extreme-noise} setting, in which we only know whether or not a node was infected. We provide a polynomial time algorithm for recovering the structure of bidirectional trees in the extreme-noise setting, and show our algorithm almost matches lower bounds established in the no-noise setting, and hence is optimal up to log-factors. We extend our results for general degree-bounded graphs, where again we show that our (poly-time) algorithm can recover the structure of the graph with optimal sample complexity. We also provide the first efficient algorithm to learn the weights of the bidirectional tree in the limited-noise setting. Finally, we give a polynomial time algorithm for learning the weights of general bounded-degree graphs in the limited-noise setting. This algorithm extends to general graphs (at the price of exponential running time), proving the problem is solvable in the general case. All our algorithms work for any noise distribution, without any restriction on the variance.
\end{abstract}

\section{Introduction}
Epidemic models accurately represent (among other processes) the spread of diseases, information  (rumors, viral videos, news stories, etc.), the spread of malevolent agents in a network (computer viruses, malicious apps, etc.), or even biological processes (pathways in cell signaling networks, chains of activation in the gene regulatory network, etc.). 

We focus on epidemics that spread on an underlying graph \cite{Newman2014a}, as opposed to the fully mixed models introduced in the early literature \cite{Bernoulli2004}.

Most settings assume we know the underlying graph and aim to study properties of the spread. Much work has been done in detection \cite{Arias-castro2011, Arias-castro, Milling2015, Milling2012, Meirom2014, Leskovec2007, Khim2017}, where the goal is to decide whether or not there is indeed an infection. This problem is of importance in deciding whether or not a computer network is under attack, for instance, or whether a product gets sold through word-of-mouth or thanks to the advertisement campaign (or both \cite{Myers2012}). More specifically, the problem of source detection \cite{Shah2010, shah2012rumor, shah2010detecting, spencer2015impossibility, wang2014rumor} or obfuscation \cite{fanti2016rumor, Fanti2014, Fanti2017} has been extensively studied. On the other side of the spectrum, both experimental and theoretical work has tackled the problem of modeling \cite{DelVicario2016, Wu2018, Gomez-Rodriguez2013}, predicting the growth \cite{Cheng2014, Zhao2015}, and controlling the spread of epidemics \cite{Drakopoulos2014, Drakopoulos2015, Hoffmann2018, Farajtabar2017}. 

In this work, we take the opposite approach: assuming we know some properties of the spread, can we recover the underlying graph? The early works on this subject proposed a few heuristics and experimentally proved their effectiveness \cite{Gomez-rodriguez2012, Iwata2013}.  Netrapalli et al. \cite{Netrapalli2012} established the first theoretical guarantees for this problem for discrete-time infections. They proved one can recover the edges of any graph with correlation decay, with access to the times of infection for multiple cascades spreading on the graph. They introduced a likelihood, proved it decouples into convex subproblems, and demonstrated that the edges of the graph can therefore be obtained efficiently. They also proved a sample complexity lower bound and showed their method is within a log factor of it. Abrahao et al. \cite{Abrahao2013} also introduced a method of solving this problem, this time for a more realistic, continuous-time infection model, through learning only the first edge of each cascade. Zarezade et al. \cite{Zarezade2017} proposed a first experimental attempt to tackle the case of correlated cascades using Hawkes processes. Khim et al. \cite{Khim2018} extended the theoretical results to the case where the the cascades spreading on the graph are not independent, which required completely new machinery involving martingales and weighted Pólya urns.

All the results above assume we have perfect knowledge of the properties of the spread we use to reconstruct the graph. For most of the literature, those are the times of infection for all nodes for each cascade. This assumption may hold for online epidemics, as information is usually dated (for instance, posts or retweets on social networks have time stamps). For human networks, however, this assumption is often unrealistic: official diagnosis (and hence recording by any tracking entity such as the CDC) may come days, weeks, or in important examples such as HIV, {\em years} after the actual moment of infection. Moreover, this can be highly variable from person to person, hence the infector is often diagnosed after the infectee. Similar issues arise with biological networks: we only know the expression of a gene when a measure is taken, which can happen after a typically arbitrary delay. 

We therefore develop a method for learning the graph of epidemics with noisy times of infection. We demonstrate that past approaches are unusable, due to the fact that even small levels of noise are typically enough to cause order-of-diagnosis to differ from order-of-infection. We develop new techniques to deal with this setting, and prove our algorithm can reliably learn the edges of a tree in the limited-noise setting, for \textit{any} noise distribution. We also show we can learn the structure of any bounded degree graph from a very weak observation model, in an sample-optimal fashion (up to log-factors). We finally provide an algorithm which learns the weights of any bounded-degree graph in the limited-noise setting.

\begin{table*}
  \caption{Notations}
  \label{notation-table}
  \begin{tabular}{cl}
    \hline
$G = (V,E)$ & Graph $G$, $V$ set of nodes, $E$ set of edges. \\
$N$  & Number of nodes in the graph. \\
$T_i^m$ & Random variable for the actual time of infection of node $i$ during cascade $m$. \\
$n_i^m$ & Noise of node $i$ during cascade $m$. \\
$T_i^{'m} = T_i^m + n_i^m $ & Random variable for the noisy time of infection of node $i$ during cascade $m$. \\
$I_i^{m} $ & Random boolean variable. ($I_i^{'m} = True \Leftrightarrow$ node $i$ was infected during cascade $m$). \\
$p_{ij}$ & Weight of edge $(i,j)$, corresponding to the probability that $i$ infects $j$.  \\
    \hline
  \end{tabular}
\end{table*}
\subsection{Model}

\begin{figure*}
\centering
\begin{subfigure}[t]{0.21\linewidth}
\centering
\includegraphics[width = \linewidth]{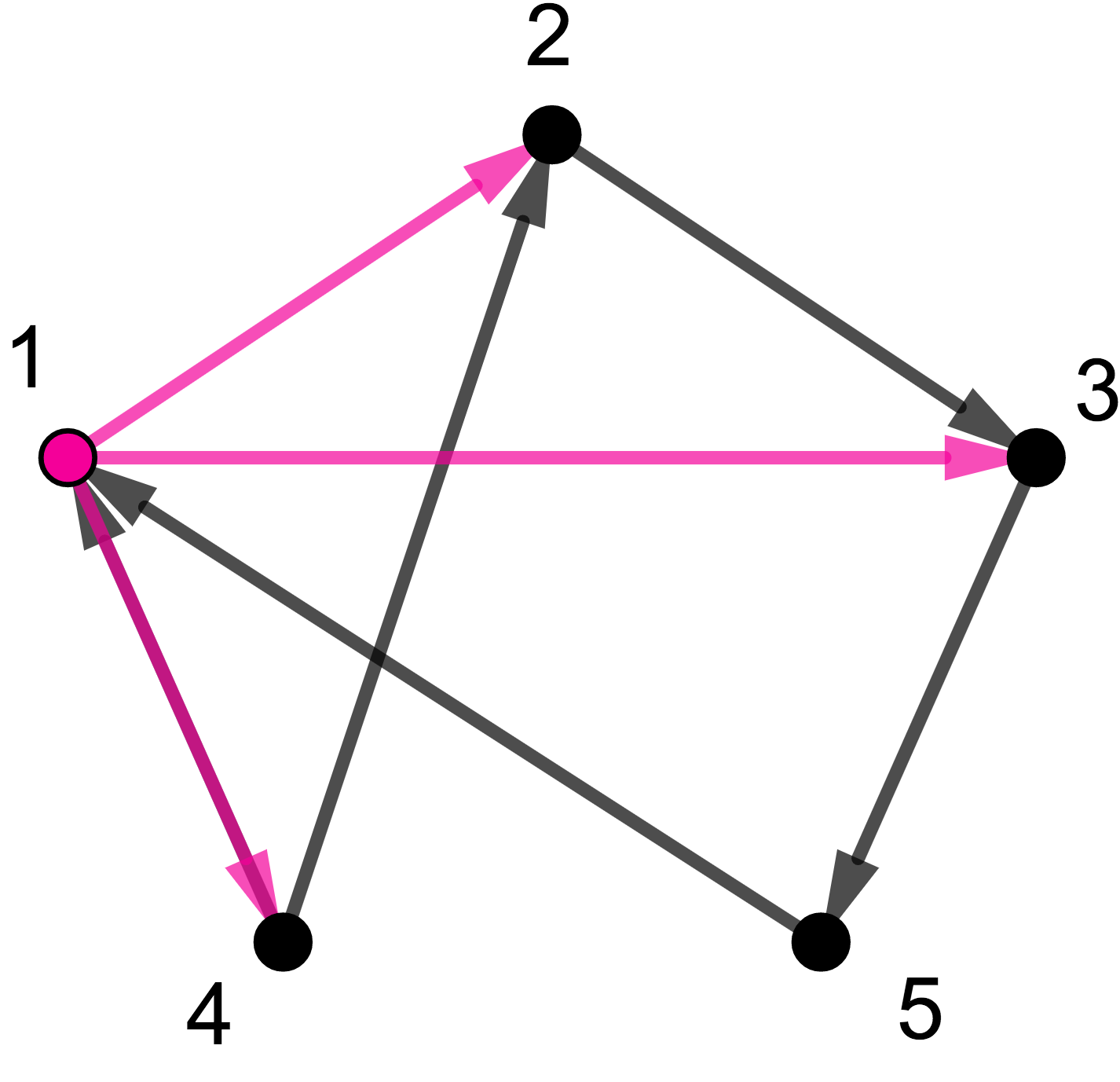}%
\caption{At t=0, node 1 is the source, in the infected state. It can possibly infect node 2, 3 and 4, all in the susceptible state.}
\end{subfigure}\qquad
\begin{subfigure}[t]{0.21\linewidth}
\centering
\includegraphics[width = \linewidth]{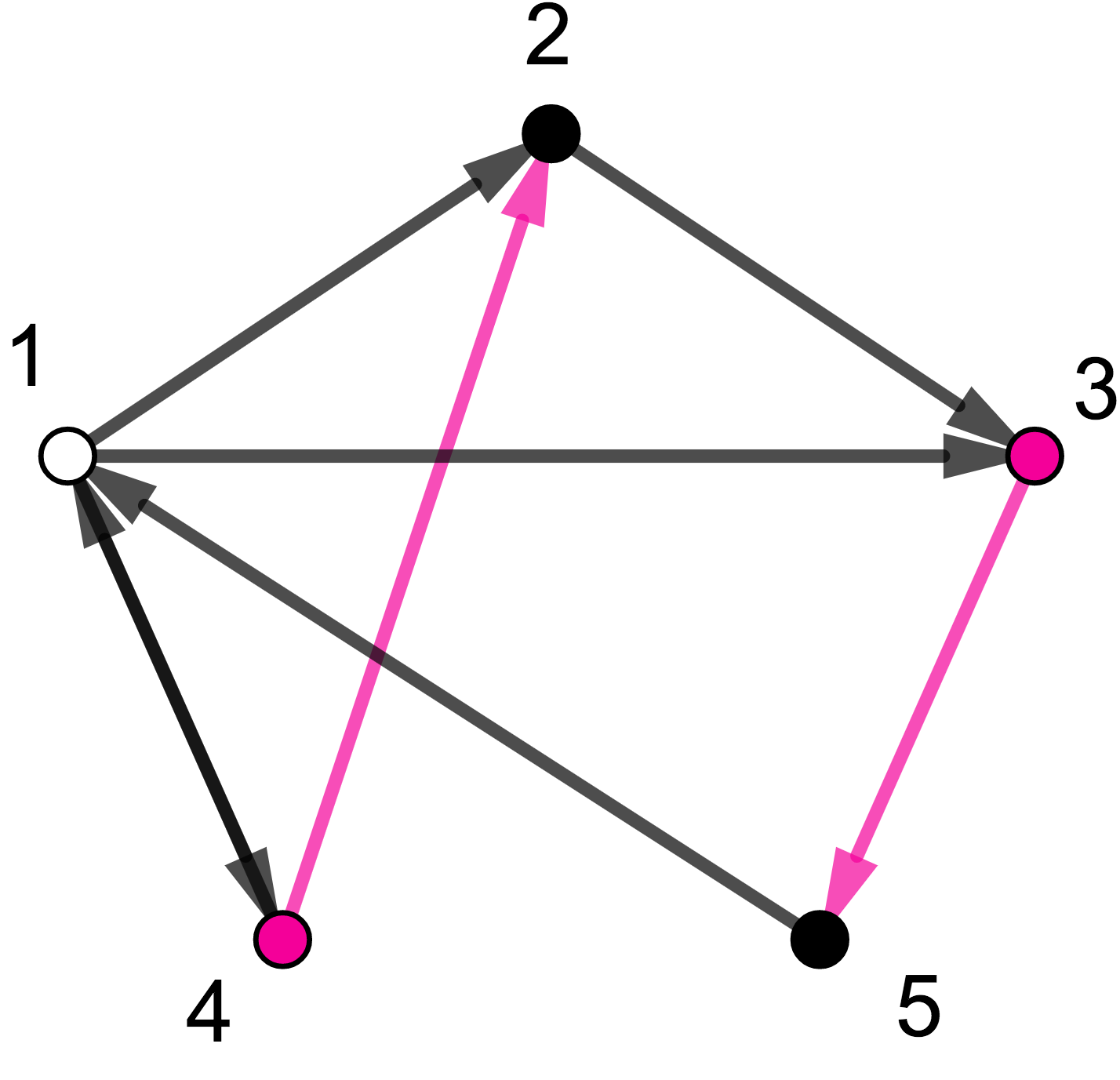}%
\caption{At t=1, nodes 3 and 4 are infected. Node 3 can infect node 5, in the susceptible state. Node 4 can infect node 2, in the susceptible state, but not node 1, since node 1 is now removed.}
\end{subfigure}\qquad
\begin{subfigure}[t]{0.21\linewidth}
\centering
\includegraphics[width = \linewidth]{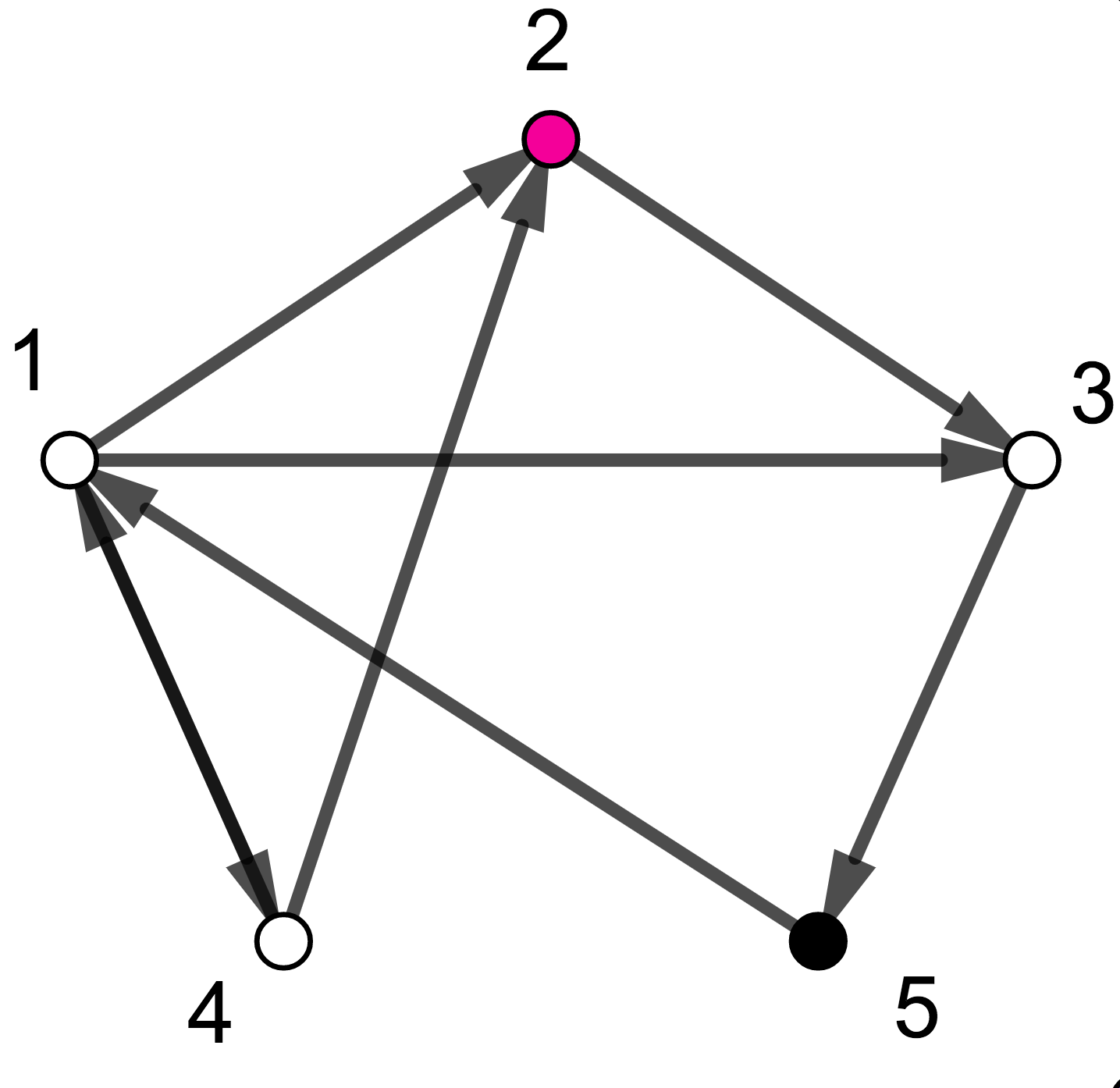}%
\caption{At t=2, node 2 is infected. All its neighbors are in the removed state, so new node can be infected.}
\end{subfigure}\qquad
\begin{subfigure}[t]{0.21\linewidth}
\centering
\includegraphics[width = \linewidth]{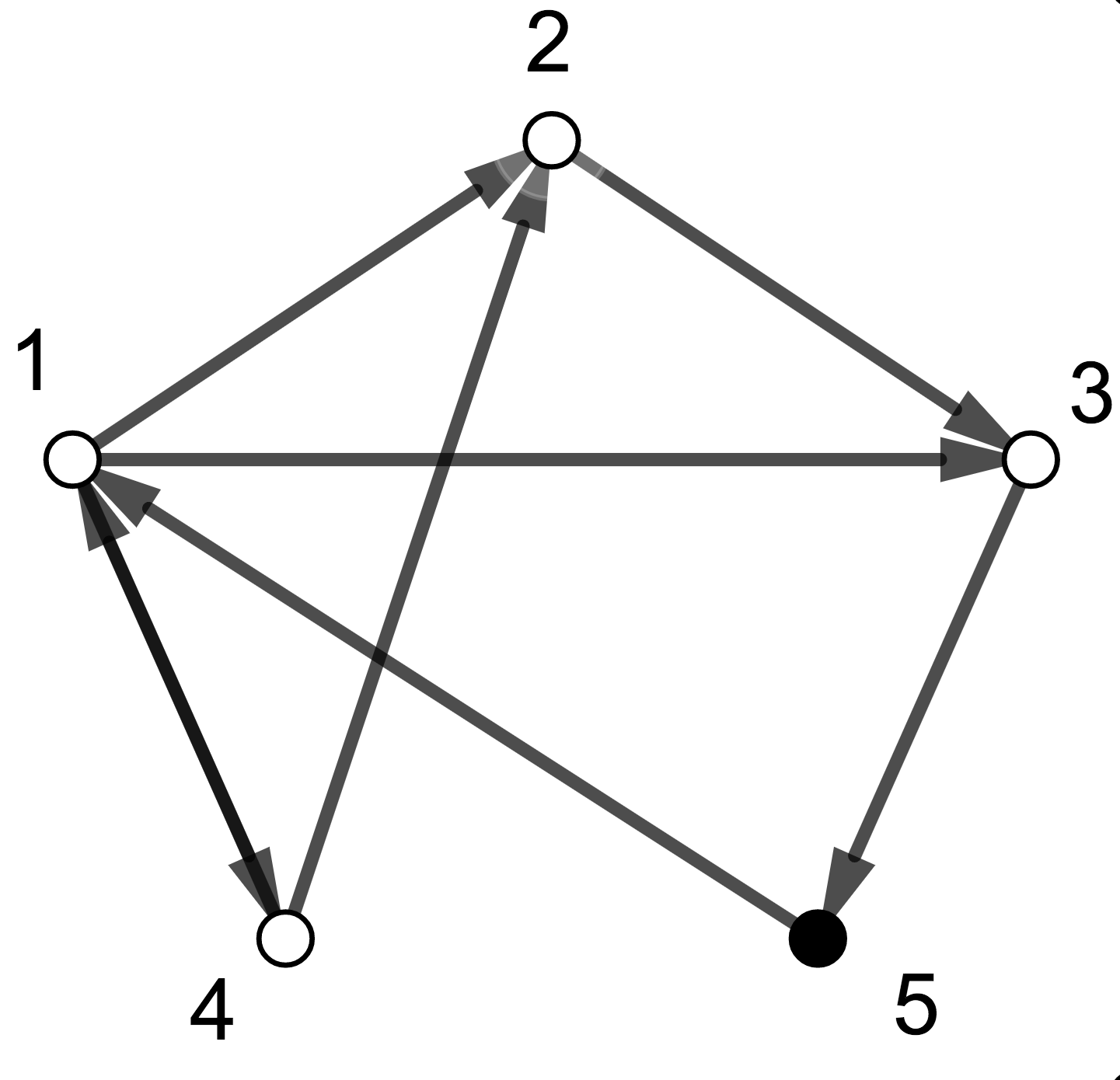}%
\caption{At t=3, the cascade stops, even if node 5 remains in the susceptible state.}
\end{subfigure}
\caption{A complete cascade.}\label{fig:spread}
\end{figure*}
We observe epidemics spreading on a graph, and aim to reconstruct the graph structure from noisy estimates of the times of infection. In this section, we specify the exact propagation model, the noisy observation model, and the two learning tasks this work tackles.

\textbf{Propagation model:} We consider a particular variant of the \textit{independent cascade model}, close to the one-step model introduced by \cite{Goldberg2001} and further studied by \cite{Kempe2003}. The epidemic spreads in \textit{discrete time} on a \textit{weighted directed graph} $G = (V,E)$, where parents can infect their children with probability equal to the weight of the edge between them, but children cannot infect their parents. We allow bidirectional edges: it is possible that both $(i,j) \in E$ and $(j,i) \in E$, possibly with different weights. For each edge $(i,j) \in E$, the corresponding weight $p_{ij}$ is such that $0<p_{min}\leq p_{ij} \leq p_{max} <1$. 

This process is an instance of a $Susceptible \rightarrow Infected \rightarrow Removed$ (SIR) process. Each node starts out in the susceptible state. As in \cite{Abrahao2013}, each cascade $m$ starts at a positive time \footnote{Most of the literature considers the initial time of infection to be 0. This is because when we have access to the exact times of infection, we can make this assumption without loss of generality. In our case, it would imply we know exactly when an outbreak started,  which is usually not the case.} on a unique node source node, picked uniformly among the nodes of the graph. Once the source becomes infected, it is removed from the graph, and if it has children, each is infected at the next time step independently according to a probability specified by the weight of the edge shared with the source. 

The process ends when there are no newly infected nodes (either because no infection happened during the previous time step, in which case some nodes may never be infected, or because all the nodes of the graph are removed). One realization of this process from start to finish is called a \textbf{cascade}. If two nodes are infected during the same cascade, we say that they are \textbf{co-infected} for this cascade. This process is illustrated in Figure \ref{fig:spread}. 

\textbf{Observation model:} Let $T_i^m$ be a random variable corresponding to the time of infection of node $i$ during cascade $m$, and let $t_i^m$ be its realization (if $i$ stays in the susceptible state during cascade $m$, we have $t_i^m = \infty$). We introduce three observation models. 

In the \textbf{no-noise setting}, we have access to the exact times of infection $T_i^m$. 

In the \textbf{limited-noise setting}, we never get to observe the exact times of infection $T_i^m$, but only a noisy version $T^{'m}_i = T^m_i + n_i^m$ (with realization $t_i^{'m}$), where all the $n_i^m$ are i.i.d., and represent the noise added to the $T^m_i$. We assume $n_i^m $ follows a \textit{known} distribution $\mathcal{D}$. The only restriction we put on $\mathcal{D}$ is that it cannot have infinite value (\textit{i.e.,} $t^{'m}_i = \infty \Leftrightarrow t^{m}_i = \infty$, and we know for a fact when nodes have been infected or not). 

In the \textbf{extreme-noise setting}, we take the previous setting to the extreme, and we assume that instead of having access to the noisy times of infection $T_i^{'m}$, we only have access to the infection status of the nodes $I_i^m$. We know $I_i^m = True$ if $i$ was infected during cascade $m$, and $I_i^m = False$ otherwise.  Note that $T_i^{'m} < \infty \Leftrightarrow I_i^m = True$, so we can always deduce the infection status from the noisy times of infection. However, we cannot guess the noisy times of infection from the infection status: the (noisy) times of infections contain strictly more information than the infection status.

For these three settings, we call a \textbf{sample} the vector of all observations for the cascade $m$. In the no-noise setting, this is the extended-integer vector $\{ t_i^{m}\}_{i\in V}$. In the limited-noise setting, this is the extended-integer vector $\{ t_i^{'m}\}_{i\in V}$. In the extreme-noise setting, this is the boolean vector corresponding to the realization of $\{ I_i^{m}\}_{i\in V}$. We also use the notation $T' = \{ T_i^{'m}\}_{i\in V}^{m = 1\dots M} $ (respectively $t' = \{ t_i^{'m}\}_{i\in V}^{m = 1\dots M}$) for the matrix representing the random variable (respectively the realizations) of all the samples.

\textbf{Learning tasks:} We focus on two different learning tasks. When we learn the \textbf{structure} of a graph, it means that for any two nodes $i$ and $j$, we can decide whether or not there exists an edge between these two nodes (whatever its direction). When we learn the \textbf{weights} of the graph, it means that for every two nodes $i$ and $j$, we learn the exact value\footnote{When $(i,j) \notin E$, we have $p_{ij} = 0$.} of both $p_{ij}$ and $p_{ji}$ up to precision $\epsilon$.

\subsection{Why is it a hard problem?} \label{sec:hard}

\subsubsection{Counting approaches}

Most approaches in the no-noise setting relate to counting. In our setting, for instance, a natural (and consistent) estimator for $p_{ij}$ is to count how often an infection occurred along an edge, and divide it by how often such an infection could have happened:
$$ \hat{p_{ij}} = \frac{\text{Number of times $j$ becomes infected one time step after $i$}}{\text{Number of times $i$ was infected before $j$}}. $$

\begin{figure}
\centering
\includegraphics[width=0.20\linewidth]{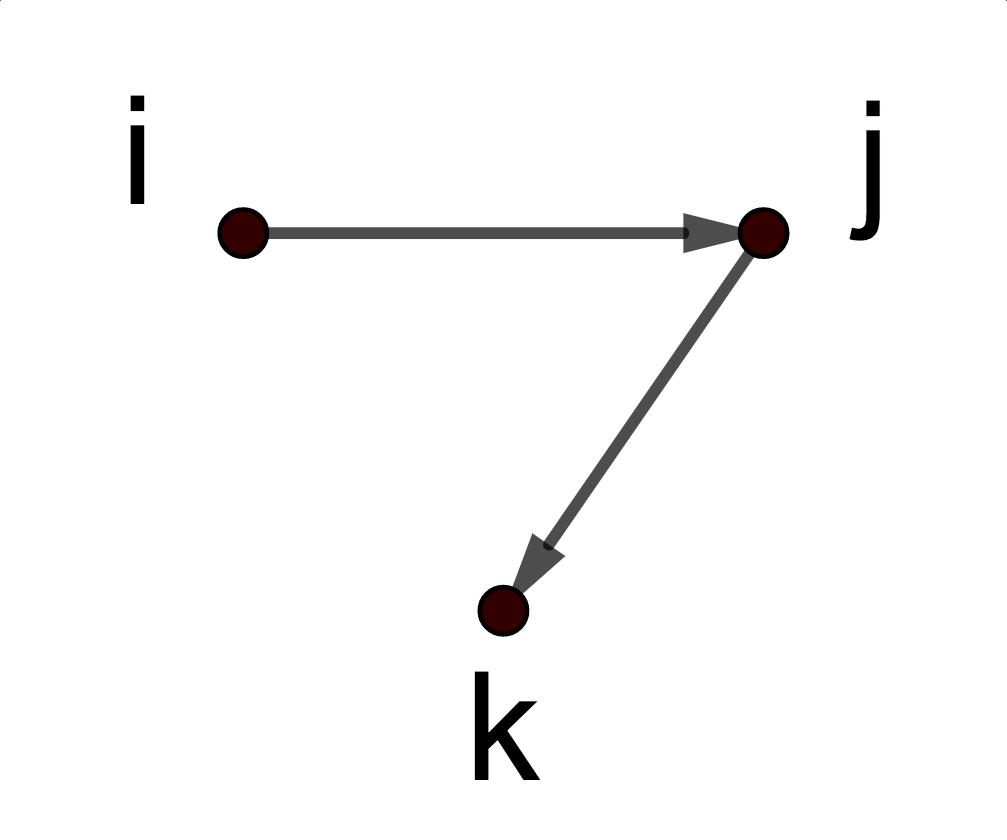}
\caption{Possible scenarios which could have led to $T^{'}_i = 2$, $T^{'}_j = 3$ and $T^{'}_k = 4$. In the no-noise setting, this implies $T_i = 2, T_j = 3, T_k=4$, and there is only one possible infection pattern.}
\end{figure}

\begin{figure}
\centering
\begin{subfigure}[t]{0.20\linewidth}
	\centering
	\includegraphics[width = \linewidth]{ijk.png}%
	\caption{}{{\begin{tabular}{c|c|c}
			& T & n \\
			\hline 
			i & 0&2 \\
			j & 1&2 \\
			k & 2& 2
	\end{tabular}}}
\end{subfigure}
\begin{subfigure}[t]{0.20\linewidth}
\centering
\includegraphics[width = \linewidth]{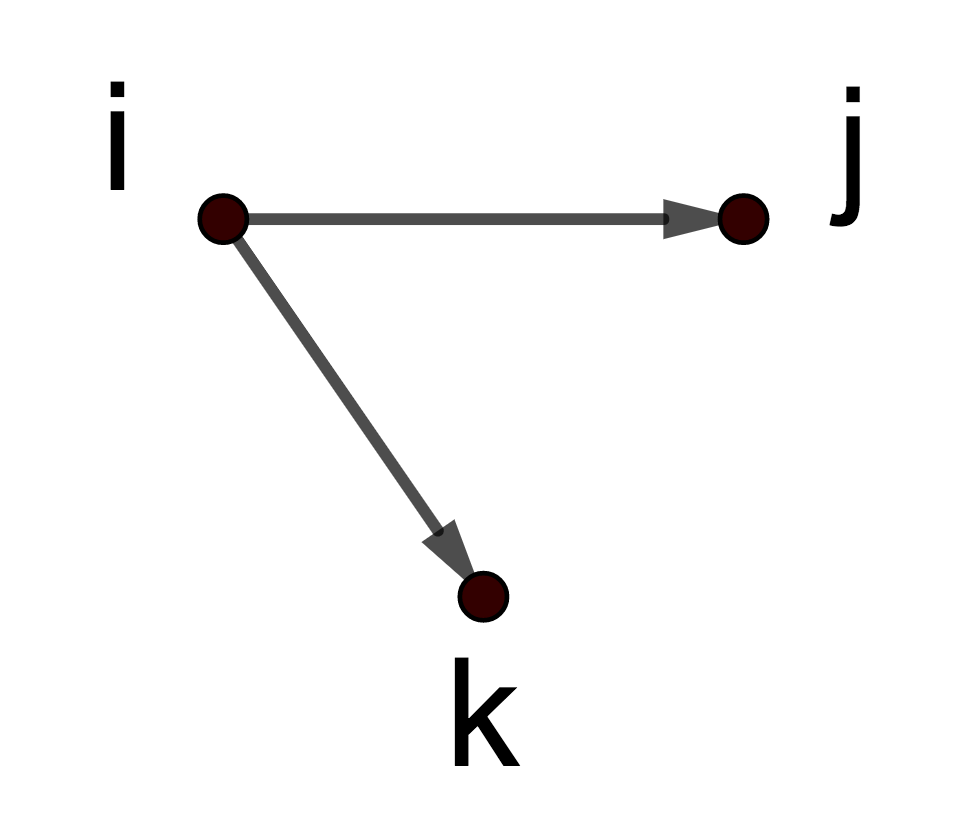}%
\caption{}{{\begin{tabular}{c|c|c}
			& T & n \\
			\hline 
			i & 0&2 \\
			j & 1&2 \\
			k & 1& 3
	\end{tabular}}}
\end{subfigure}
\begin{subfigure}[t]{0.20\linewidth}
\centering
\includegraphics[width = \linewidth]{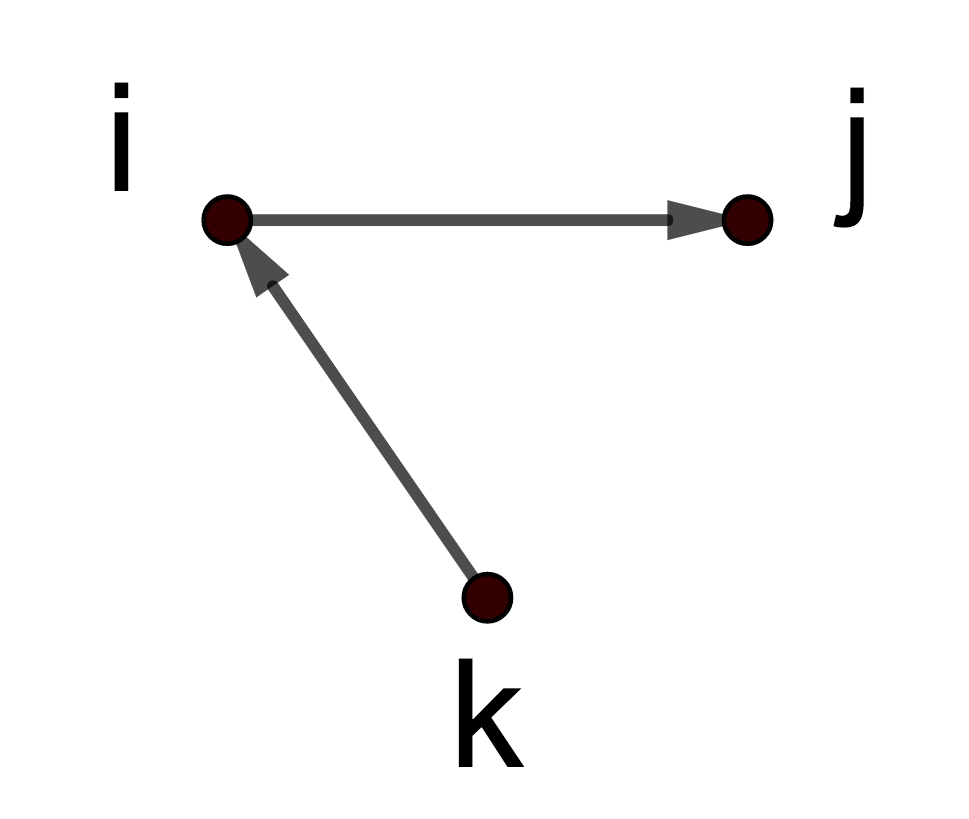}%
\caption{}{{\begin{tabular}{c|c|c}
         & T & n \\
         \hline 
        i & 1&1 \\
        j & 2& 1\\
        k & 0& 4
	\end{tabular}}}
\end{subfigure}

\begin{subfigure}[t]{0.20\linewidth}
\centering
\includegraphics[width = \linewidth]{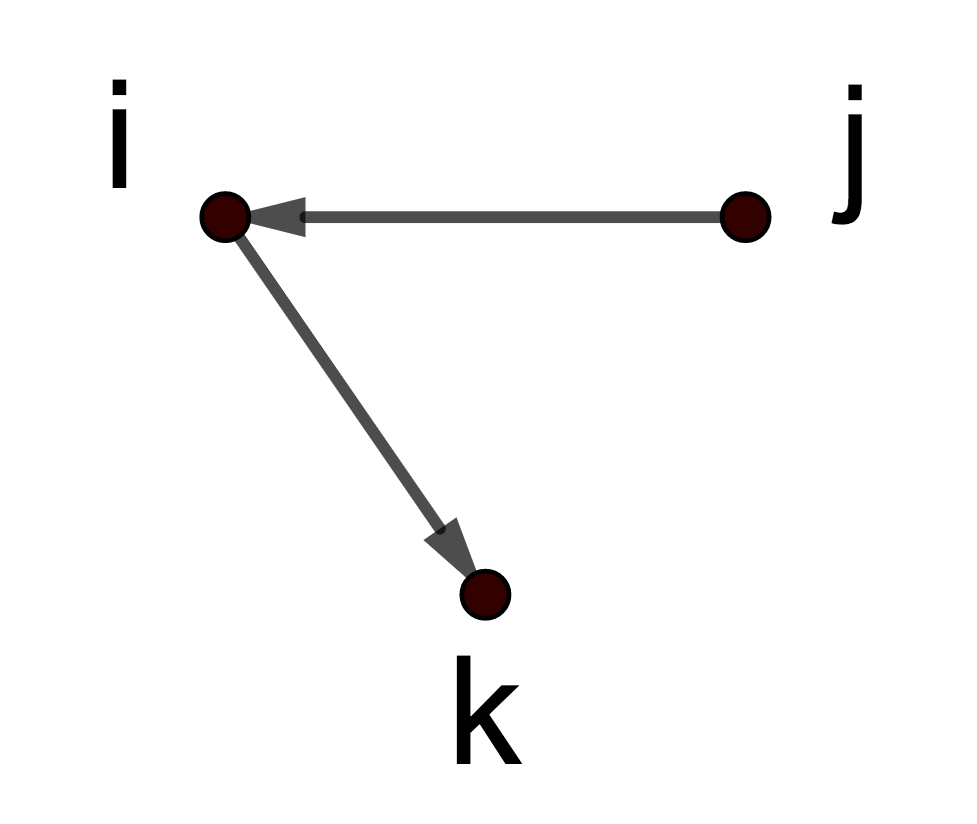}%
\caption{}{{\begin{tabular}{c|c|c}
         & T & n \\
         \hline 
        i & 1&1 \\
        j & 0&3 \\
        k & 2& 2
	\end{tabular}}}
\end{subfigure}
\begin{subfigure}[t]{0.20\linewidth}
\centering
\includegraphics[width = \linewidth]{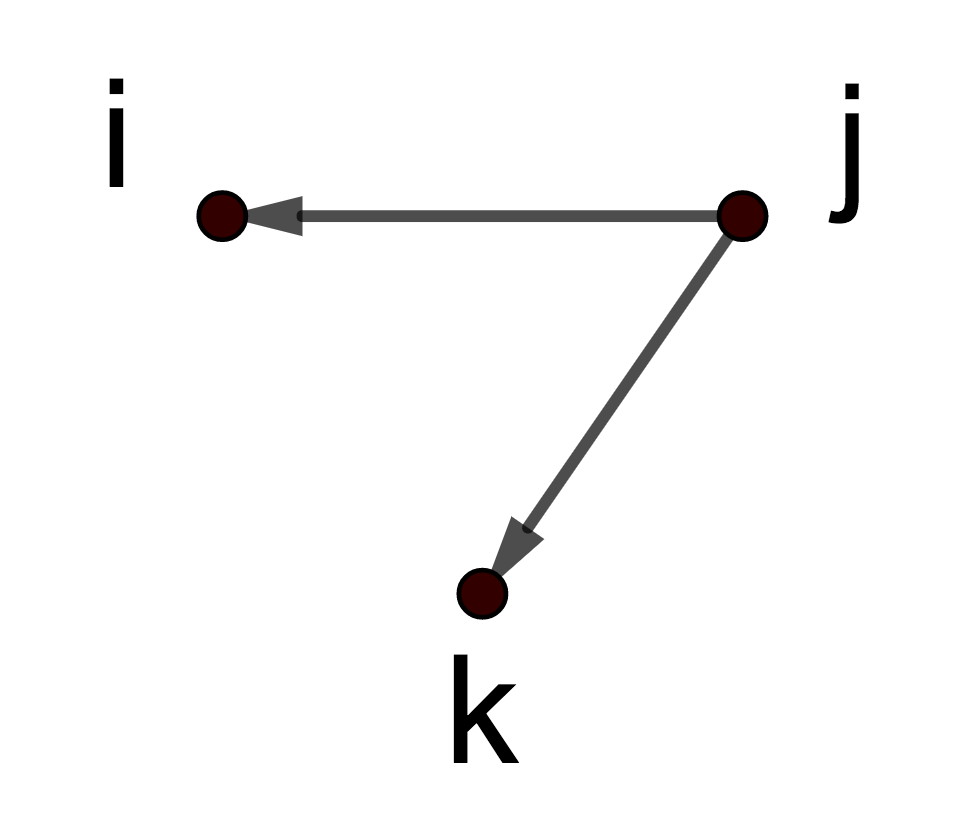}%
\caption{}{{\begin{tabular}{c|c|c}
         & T & n \\
         \hline
        i & 1&1 \\
        j & 0& 3\\
        k & 1& 3
	\end{tabular}}}
\end{subfigure}
\begin{subfigure}[t]{0.20\linewidth}
\centering
\includegraphics[width = \linewidth]{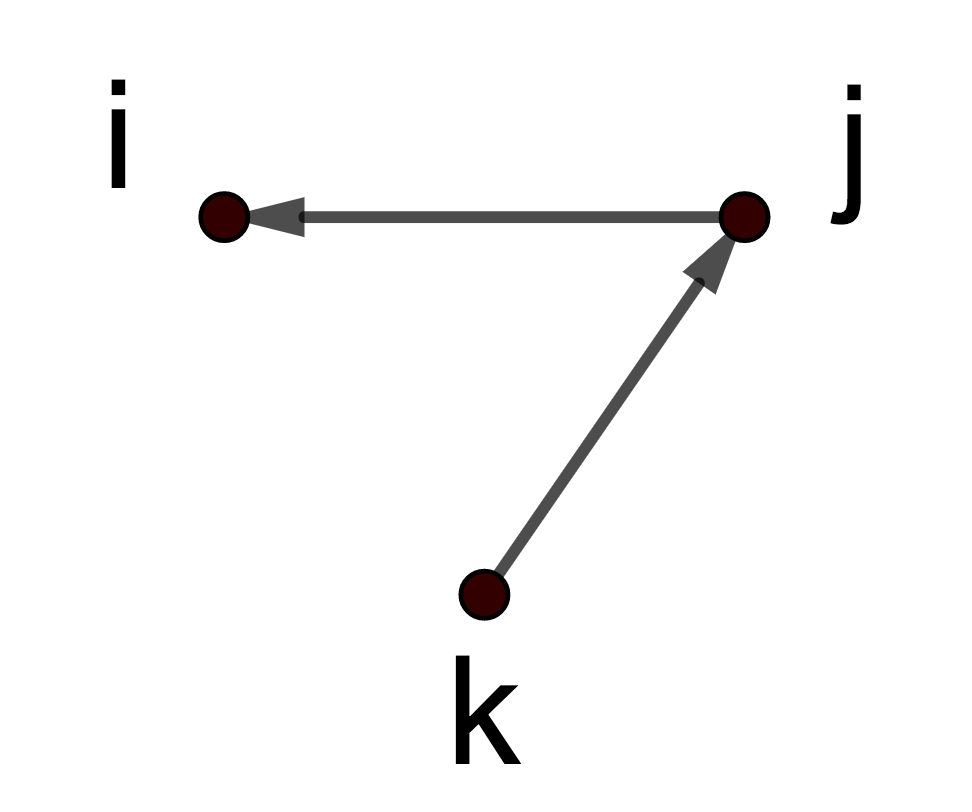}%
\caption{}{{\begin{tabular}{c|c|c}
         & T & n \\
         \hline
        i & 2&0 \\
        j & 1&2 \\
        k & 0& 4
	\end{tabular}}}
\end{subfigure}

\begin{subfigure}[t]{0.20\linewidth}
\centering
\includegraphics[width = \linewidth]{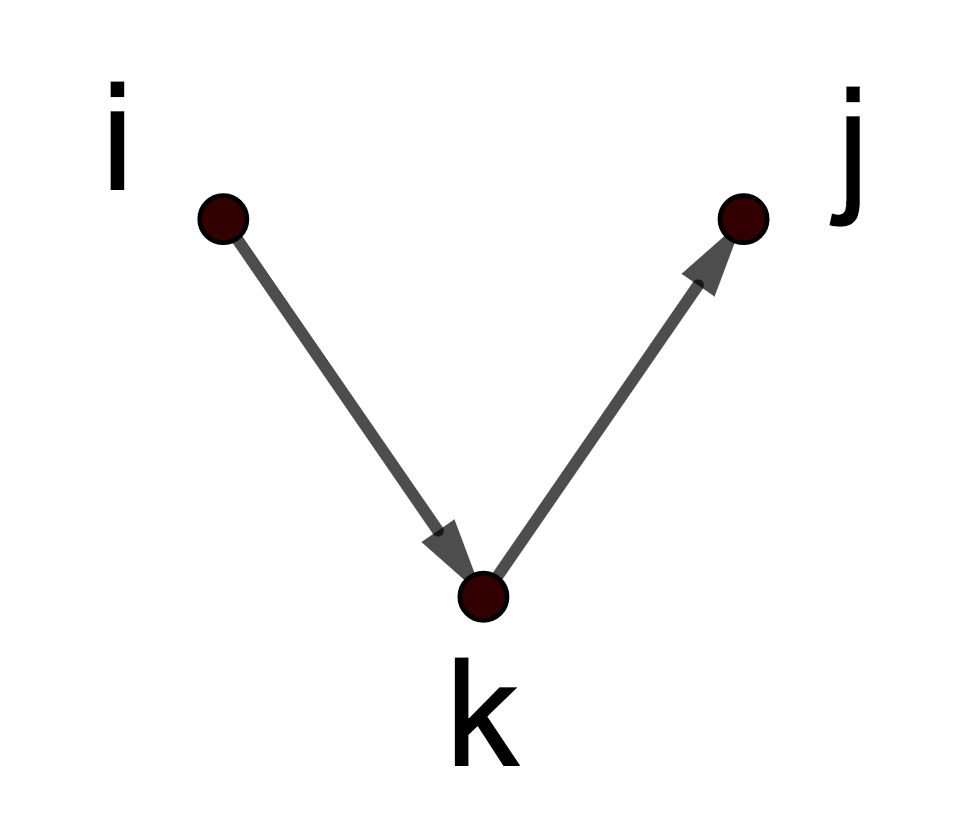}%
\caption{}{{\begin{tabular}{c|c|c}
         & T & n \\
         \hline
        i & 0&2 \\
        j & 2&1 \\
        k & 1&3 
	\end{tabular}}}
\end{subfigure}
\begin{subfigure}[t]{0.20\linewidth}
\centering
\includegraphics[width = \linewidth]{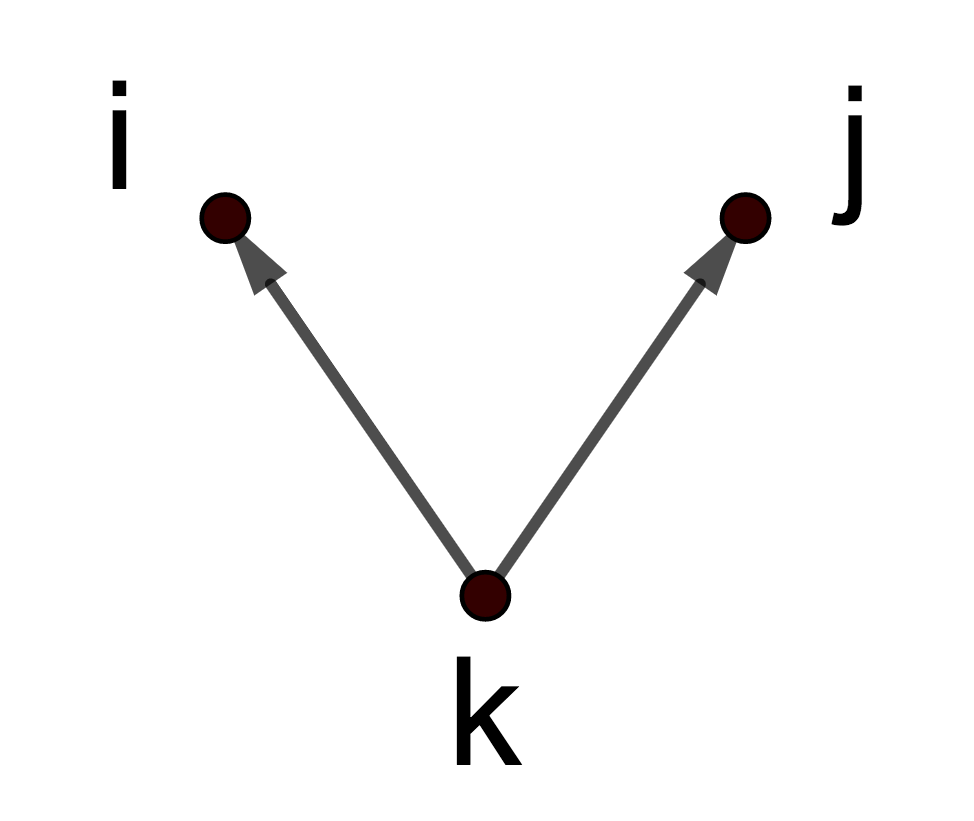}%
\caption{}{{\begin{tabular}{c|c|c}
         & T & n \\
         \hline
        i & 1&1 \\
        j & 1&2 \\
        k & 0&4 
	\end{tabular}}}
\end{subfigure}
\begin{subfigure}[t]{0.20\linewidth}
\centering
\includegraphics[width = \linewidth]{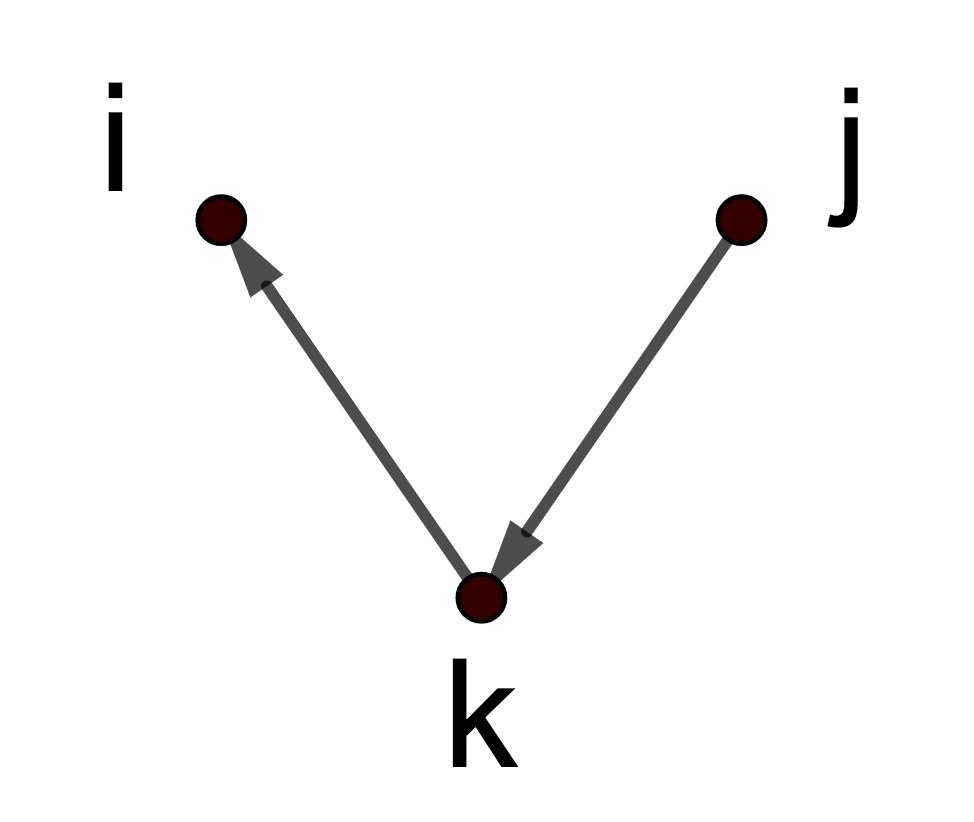}%
\caption{}{{\begin{tabular}{c|c|c}
         & T & n \\
         \hline
        i & 2&0 \\
        j & 0&3 \\
        k & 1&3 
	\end{tabular}}}
\end{subfigure}
\caption{Possible scenarios which could have led to $T^{'}_i = 2$, $T^{'}_j = 3$ and $T^{'}_k = 4$. We have $T^{'}_l = T_l + n_l$. In the limited-noise setting, there are nine possible infection patterns (many more scenarios with the same infection pattern, but different noise values, are not shown).}
\end{figure}

In the no-noise setting, $j$ could only have been infected by a node signaling exactly one time step before $j$. However, in the limited-noise setting, $j$ signaling its infection one time step after $i$ could stem from a variety of scenarios:
\begin{itemize}
    \item $i$ could have indeed infected $j$: cases a), b) and c) of Figure \ref{fig:second} above.
    \item $j$ could have infected $i$, but the noise flipped the order of signaling: cases d), e) and f).
    \item No infection happen between $i$ and $j$, and the probability of infectin depends mainly on another node $k$: cases g), e) and f). This could happen for \textit{any other node $k$} in the graph.
\end{itemize}
The natural estimator introduced earlier is therefore not consistent anymore; instead, it tends to a quantity which depends on $p_{ij}$, but also $p_{ji}$, and $p_{ik}, p_{ki}, p_{jk}, p_{kj}$ as well, for all the other nodes $k$ in the graph. By counting the number of times $j$ became infected one time step after $i$, we are not counting the number of infections along the edge $(i,j)$ anymore, but instead a mixture of all the scenarios described above, which not only include the cases where $j$ infected $i$, but also events in which the cascade spread through another node $k$, and the edge $(i,j)$ was irrelevant to the process. Using this estimator, or any obvious (to us) extension of it, would not only imply learning the wrong weights for the edges, but also learning edges when there are no edges. Our first contribution is therefore to design a new set of estimators, from which we can deduce the value of $p_{ij}$ (Sections \ref{sec:treeWeights} and \ref{sec:cascade2}).

Adding noise in the time of infection not only reverses the cascade chronology, it also exponentially increased the number of possible infection patterns that \textit{could} have happened. Bounding the realm of possibilities is therefore our second step towards solving the problem (Section \ref{sec:treeStruct}).


\subsubsection{Max-likelihood approaches}
Another common approach is to use likelihood-based methods. For instance, in \cite{Netrapalli2012}, the authors develop a max-likelihood-based approach to learn the edges of the graph. They prove the log-likelihood has desirable properties: it decouples into only one local problem for each node, and this local problem is convex (up to the change of variable $\theta_{ij}=-\log(1 - p_{ij})$):
\begin{align*}
\mathcal{L}(T, P_{ij}) &= \log \left(\frac{1}{N}\cdot \prod_{1 \leq i \leq N} \underbrace{\left(\prod_{j; t_j < t_i - 1} (1 - p_{ji}) \right)}_{\substack{\text{Probability that $i$ was not} \\ \text{infected before $t'_i$.}}}\cdot\underbrace{\left(1 - \prod_{j; t_j = t_i - 1} (1 - p_{ji}) \right)}_{\substack{\text{Probability that $i$ was} \\ \text{infected at $t'_i$.}}} \right) \\
\mathcal{L}(T, P_{ij}) &= \log\left(\frac{1}{N}\right) + 
\sum_{i = 1}^{N} \sum_{j; t_j < t_i - 1} \log(1 - p_{ji})+ \sum_{j; t'_j = t'_i - 1} \log(1 - p_{ji})
\end{align*} 

In our setting, the log-likelihood has none of these properties. It is not convex, and it is unclear any method other than brute force could find its maximum. Moreover, it does not decouple anymore, and even computing the log-likelihood itself takes exponential time. 

\begin{align*}
\mathcal{L}(T, P_{ij}) &= \log\left(\frac{1}{N}\cdot \underbrace{\sum_{(t_1, \dots, t_N) \leq (t'_1 - 1, \dots, t'_N - 1)}}_{\substack{\text{This is the root} \\ \text{of the difficulty.}}} \cdot\prod_{1 \leq i \leq N} \underbrace{\Bigg(n(t'_i - t_i)\Bigg)}_{\substack{\text{Noise at node} \\ \text{ $i$ is $t'_i - t_i$.}}} \right. \\
&\qquad \quad \cdot \left.\underbrace{\left(\prod_{j; t_j < t_i - 1} (1 - p_{ji}) \right)}_{\substack{\text{Probability that $i$ was not} \\ \text{infected before $t'_i$.}}} \cdot \underbrace{\left(1 - \prod_{j; t_j = t_i - 1} (1 - p_{ji}) \right)}_{\substack{\text{Probability that $i$ was} \\ \text{infected at $t'_i$.}}} \right) \\
&= \log\left(\frac{1}{N} \right) + \log\left(\sum_{\substack{(t_1, \dots, t_N) \leq  \\ (t'_1 - 1, \dots, t'_N - 1)}} \prod_{1 \leq i \leq N} n(t'_i - t_i) \left(\prod_{j; t_j < t_i - 1} (1 - p_{ji}) \right)\cdot\left(1 - \prod_{j; t_j = t_i - 1} (1 - p_{ji}) \right)\right)
\end{align*} 


When dealing with hidden variables, a common technique would be to use the Expectation-Maximization algorithm \cite{Dempster1977}. However, in our setting, the number of hidden states is $\displaystyle\prod_{i = 1}^{N} t'_i$, which can be as large as $(N-1)^N$. This prohibits any realistic use of the Expectation-Maximization algorithm for networks with more than twelve nodes. Moreover, except for the recent contributions \cite{Kwon2019}, very little is known about the theoretical convergence of the Expectation-Maximization algorithm.

\subsection{Contributions}
The contributions of this article are multiple:
\begin{itemize}
    \item To the best of our knowledge, we are the first to tackle the problem of learning the edges of a graph from noisy times of infection, a simple but natural extension of a well-known problem.
    \item We provide the first efficient algorithm for learning the structure and weights of a bidirectional tree in this setting. We also establish a tree-specific lower bound which shows that our algorithm is sample-optimal (up to log-factors) for learning the structure of the tree (Section \ref{sec:tree}).
    \item We prove it is possible to learn the structure of any bounded-degree graph in the extreme setting for which we only have access to the infection status (\textit{i.e.,} whether or not a node was infected). Moreover, we can do so with almost optimal sample complexity, according to the bound established in \cite{Netrapalli2012}.
    \item We provide polynomial algorithms for learning the weights of bounded degree graphs.
    \item Finally, we extend the results from bounded-degree graphs to general graphs. This proves the problem is solvable under any noise distribution, although the exponential sample complexity and running time prohibits any use of this algorithm in practice (Section \ref{sec:cascade2}).
\end{itemize}
\section{Learning bidirectional trees} \label{sec:tree}
The bidirectional tree is the simplest example which illustrates some of the difficulties of the noisy setting. Indeed, for a directed tree, the true sequence of infections can be reconstructed, and we can use techniques from the no-noise setting. For a bidirectional tree, those techniques cannot be extended. However, the uniqueness of paths in the bidirectional tree still makes this problem considerably easier than the general setting. We therefore start by presenting a solution for the bidirectional tree. The key ideas here generalize to the neighborhood-based decomposition we introduce below, which forms our key conceptual approach for the general problem. 

This section contains three contributions. First, we show how to learn the structure of a tree using only the infection status, {\em i.e.}, what we call the extreme-noise setting (Section \ref{sec:treeStruct}). For each cascade, we only know which nodes were infected. We show this contains enough information to learn the structure of bidirectional tree. Second, we establish a lower bound for the no-noise setting, and show our algorithm has the same dependency in the number of nodes $N$ as this lower bound (up to log-factors). In other words, for the task of learning the structure of any tree, an optimal algorithm in the no-noise setting would need as many cascades as our algorithm needs in the extreme-noise setting (up to log-factors).

Finally, we show how we can leverage this learned structure to learn the weights of the tree, this time when we have noisy access to the times of infection, {\em i.e.,} the limited-noise setting (Section \ref{sec:treeWeights}). We provide sample complexity for this task.

\subsection{Tree structure} \label{sec:treeStruct}
As illustrated in Section \ref{sec:hard}, the number of edges that \textit{could} exist is much higher in the limited-noise setting than the number of actual edges in the tree. Our first key contribution is therefore to introduce a new estimator, $\hat{h}_{ij}$, which keeps track of the fraction of cascades for which $i$ and $j$ were both infected. This estimator can therefore be computed only with the infection status in the extreme-noise setting. Using this estimator, we show that in the specific case where the graph is a tree, we learn the structure of this tree, $i.e.$ whether or not both $p_{ij} = p_{ij} = 0$.

Our algorithm for learning the edges of the tree relies on one central observation: $h_{i,j}$ achieves a kind of local maximum if there is an edge between $i$ and $j$ (Lemma \ref{cl:decrease}). This observation relies heavily on the fact that there is uniqueness of paths on a tree. Let us now dive into the proof.

\begin{definition}
 Let $\hat{h}_{i,j}$ be the fraction of cascades in which both $i$ and $j$ became infected. We have:
$$ \hat{h}_{i,j} \to_{M \to\infty} \p(I^{m}_i \,\,\& \,\, I^{m}_j ) := h_{i,j}.$$
\end{definition}

We now show that the limit $h_{i,j}$ of the estimator $\hat{h}_{i,j}$ satisfies a local maximum property on the edges of the tree:
\begin{lemma} \label{cl:decrease}
If $i$ and $j$ are not neighbors, let $(u_0, u_1, \dots, u_d)$ be the path between them, with $u_0 =i$, $u_d = j$, and $d>1$. Then:
$$\forall r \in \{0, d-1\}, \,  h_{i,j} < h_{u_r, u{r+1}}.$$
\proof 
We consider the case in which both $i$ and $j$ have been infected. There is a unique source of infection and a unique path between $i$ and $j$. Therefore, all the nodes on the path from $i$ to $j$ must have been infected as well. In particular, both $u_r$ and $u_{r+1}$ were infected.  This shows $I^{m}_i  \,\,\&\,\,  I^{m}_j  \Rightarrow  I^{m}_{u_r} \,\,\&\,\,  I^{m}_{u_{r+1}} $, so  $\p(I^{m}_i \,\,\& \,\, I^{m}_j ) \leq \p(I^{m}_{u_{r}} \,\,\& \,\,  I^{m}_{u_{r+1}})$, and therefore  $  h_{i,j} \leq h_{u_r, u_{r+1}} $.

What's more, every time $u_r$ and $u_{r+1}$ became infected, at least one more infection along an edge must have occurred in order for $j$ to become infected as well. This occurred with probability at most $p_{max}<1$. Therefore, $ h_{i,j} =  \p(I^{m}_i \,\,\& \,\,  I^{m}_j ) \leq p_{max}\cdot \p(I^{m}_{u_{r}} \,\,\& \,\,  I^{m}_{u_{r+1}} ) = p_{max}\cdot h_{u_r, u_{r+1}} $. We conclude $h_{i,j} < h_{u_r, u_{r+1}}$. 
\\\qed
\end{lemma}

This simple lemma allows us to design Algorithm \ref{alg:learnEdges}. Indeed, suppose we have access to all the limits $h_{i,j}$. By ordering them in decreasing order, we can deduce the structure of the tree by greedily adding every edge unless it forms a cycle\footnote{This algorithm is very similar in spirit to Kruskal's algorithm for finding the maximum spanning tree \cite{}}. 

\begin{algorithm}[H]
\caption{Learn the undirected edges of the tree.}\label{alg:learnEdges}
\begin{algorithmic}[1]
\Procedure{LearnTree}{$\{h_{i,j} \}_{i,j \in V}$} \Comment{$h_{i,j}$ limit of $\hat{h}_{i,j}$.}
\State $pairs\_h\_edge \gets [(h_{i,j}, (i, j))$ for $1\leq i \leq j \leq n_{nodes} ]$
\State $\textsc{Sort}(pairs\_h\_edge)$ by decreasing order 
\State $edges\_tree \gets [] $
\For{$ (\sim \, , potential\_edge) \in pairs\_h\_edge$} \label{line:11}
\If{Adding $potential\_edge$ to $edges\_tree$ does not create a cycle}
\State Add $potential\_edge$ to $edges\_tree$
\EndIf
\EndFor
\Return $edges\_tree$
\EndProcedure
\end{algorithmic}
\end{algorithm}
We show that if we have access to the limits $h_{i,j}$ of the estimators $\hat{h}_{i,j}$, the algorithm above correctly find the structure of the tree.

\begin{lemma}\label{lem:alg}
Algorithm \ref{alg:learnEdges} correctly finds all the $N-1$ pairs $(i,j)$ such that there exists at least one directed edge between $i$ and $j$.
\proof 
We show that in the for-loop at line \ref{line:11}, we add an edge to $edges\_tree$ if and only if this edge was a real edge in the original tree. We prove it by induction on the elements of the sorted list $pairs\_h\_edge$.

When no element has been selected, the proposition is trivially true. \\
Suppose now that $t$ elements of $pairs\_h\_edge$ have been examined so far. Let $ (\sim, (i,j))$ be the $t+1^{th}$ element. Two cases arise:
\begin{enumerate}
    \item $i$ and $j$ are not neighbors. Let $(u_0, \dots, u_d)$ be the path between them, with $u_0 =i$ and $u_d=j$. In this case, using Lemma \ref{cl:decrease}, $\forall r, \, h_{u_r, u_{r+1}} > h_{i,j}$. In other words, all the pairs $(u_r, u_{r+1})$ have already been considered by the algorithm. By induction, we have kept all of them in $edges\_tree$. Therefore, adding the pair $(i,j)$ would form a cycle. This pair is not kept in $edges\_tree$, which is what we wanted since it is not an edge in the original tree.
    \item $i$ and $j$ are neighbors. Suppose that adding this pair forms a cycle. Then there is a sequence $(v_0 = i,\dots, v_d = j)$ of nodes such that $h_{v_k, v_{k+1}}$ were all bigger than $h_{i,j}$, and the pairs $(v_k, v_{k+1})$ were kept by the algorithm for all $k$. However, by uniqueness of paths in a tree, there exists a pair $(v_a, v_{a+1})$ such that the path connecting $v_a$ and $v_{a+1}$ in the original tree goes through $(i,j)$. Using Lemma \ref{cl:decrease}, this means $h_{i,j} > h_{v_a, v_{a+1}}$, which is a contradiction. Therefore, adding this pair in $edges\_tree$ does not form a cycle. This pair is kept in $edges\_tree$.
\end{enumerate}
Therefore, this algorithm keeps all the edges, and only the edges of the tree, so it recovers the tree structure.
\\\qed
\end{lemma}
We next quantify how many cascades $M$ are needed for Algorithm \ref{alg:learnEdges} to be correct if we replace the $\{h_{i,j}\}_{i, j \in V}$ by their estimates $\{\hat{h}_{i,j}\}_{i, j \in V}$. We note that we do not require $\hat{h}_{i,j}$ to be close to their limit, but only need the order of the $\hat{h}_{i,j}$ to be the same as the order of the $h_{i,j}$. We identify events which guarantee that the order is the same (Corollary \ref{cor:alg}): 

\begin{definition}
Let $\mathcal{H}_3 := \{(i,j,k) \in \{1,\dots, N\}^3, p_{ij} + p_{ji} > 0 \,\, \& \,\, p_{jk} + p_{jk} > 0 \}$ be the set of triplets of nodes such that at least one directed edge exists between the first and the second node, as well as between the second and the third node.
\end{definition}

\begin{proposition} \label{cl:H3}
 If: 
 $$\forall (i,j,k) \in \mathcal{H}_3, \,\, \hat{h}_{i,j} > \hat{h}_{i,k} \text{ and } \hat{h}_{j,k} > \hat{h}_{i,k},$$ 
then for all paths $(u_0, \dots, u_d)$ in the tree, with $d>1$, we have: $$\forall r \in \{0, \dots, d-1\}, \,  \hat{h}_{u_r, u_{r+1}} > \hat{h}_{u_0, u_d}.$$
\proof For $r \in \{0, \dots, d-2\}$, we have by hypothesis $\hat{h}_{u_r, u_{r+1}} > \hat{h}_{u_r, u_{r+2}}$. Now, we recall that $\hat{h}_{u_0,u_d}$ is the number of cascades for which both $u_0$ and $u_d$ were infected. By uniqueness of paths in the tree, every time both $u_0$ and $u_d$ were infected, both $u_r$ and $u_{r+2}$ must have been infected as well. This shows that $\hat{h}_{u_0,u_d} \leq \hat{h}_{u_r,u_{r+2}}$. Notice that this is a deterministic property, not an asymptotic property. Therefore, $\hat{h}_{u_r, u_{r+1}} > \hat{h}_{u_r, u_{r+2}} \geq\hat{h}_{u_0, u_d}$. 

For $r = d-1$, we follow an identical reasoning, but with $\hat{h}_{u_r, u_{r+1}} > \hat{h}_{u_{r-1}, u_{r+1}}$.
\end{proposition}

\begin{corollary} \label{cor:alg}
If: $$\forall (i,j,k) \in \mathcal{H}_3, \,\, \hat{h}_{i,j} > \hat{h}_{i,k} \text{ and } \hat{h}_{j,k} > \hat{h}_{i,k},$$
then the correctness of Algorithm \ref{alg:learnEdges} is preserved when given $\hat{h}$ as input instead of h. \end{corollary}
In other words, Algorithm \ref{alg:learnEdges} outputs a correct set of undirected edges with finite samples.
\proof According to Proposition \ref{cl:H3}, for all paths $(u_0, \dots, u_d)$ in the tree, with $d>1$, we have that $\forall r \in \{0, \dots, d-1\}, \,  \hat{h}_{u_r, u_{r+1}} > \hat{h}_{u_0, u_d}$. As shown in the proof of Lemma \ref{lem:alg}, this is the only property of the input needed in order to yield the correct output.

\begin{proposition} \label{cl:finiteUndirected}
With $M = \frac{N \left(\log\left(\frac{1}{\delta}\right) + 2\log(N)\right)}{p_{min}(1-p_{max})}$ cascades, with probability at least $1 -\delta$, we have:
$$\forall (i,j,k) \in \mathcal{H}_3, \,\, \hat{h}_{i,j} > \hat{h}_{i,k} \text{ and } \hat{h}_{j,k} > \hat{h}_{i,k}.$$

\proof Let us consider one triplet $(i,j,k)$ in $\mathcal{H}_3$. We recall that $\hat{h}_{i,k}$ is the number of cascades for which both $i$ and $k$ were infected. Since the only path from $i$ to $k$ is through $j$, we always have that $\hat{h}_{i,j} \geq \hat{h}_{i,k} \text{ and } \hat{h}_{j,k} \geq \hat{h}_{i,k}$. We notice that to obtain $\hat{h}_{i,j} > \hat{h}_{i,k} $, we only need one cascade for which both $i$ and $j$ got infected, but not $k$. We lower bound the probability $P_{\text{triplet identified}}$ of this cascade happening. For each cascade $m$, we have:
\begin{align*}
    P_{\text{triplet identified}} &= \p(I^{m}_i\,\, \& \,\, I^{m}_j \,\,\&\,\, \textsc{Not}(I^{m}_k)) \\
    &\geq \p(\text{$i$ was a source, $i$ infected $j$, $j$ did not infect $k$}) \\
    &\quad+  \p(\text{$j$ was a source, $j$ infected $i$, $j$ did not infect $k$}) \\
    &\geq \frac{1}{N}\cdot p_{ij} \cdot (1-p_{jk}) + \frac{1}{N}\cdot p_{ji} \cdot (1-p_{jk}) \\
    &\geq \frac{1}{N} p_{min} (1-p_{max}).
\end{align*} 
The probability that this event never occurs during the $M$ cascades is upper bounded by:
\begin{align*}
    \p(\hat{h}_{i,j} = \hat{h}_{i,k}) &\leq \left(1 - \frac{1}{N} p_{min} (1-p_{max}) \right)^M \\
    &\leq e^{-\frac{M}{N} p_{min} (1-p_{max})} \\
    &\leq \frac{\delta}{N^2}.
\end{align*}
Now, there are $N-1$ edges in a tree, therefore $|\mathcal{H}_3| \leq (N-1)^2 < N^2$. By union bound:
\begin{align*}
    \p\left(\bigcup_{(i,j,k) \,\in \,\mathcal{H}_3} \hat{h}_{i,j} = \hat{h}_{i,k}\right) &\leq \sum_{(i,j,k) \,\in \,\mathcal{H}_3)} \p(\hat{h}_{i,j} = \hat{h}_{i,k}) \\
    &< N^2 \cdot \frac{\delta}{N^2} \\
    &\leq \delta.
\end{align*}
Notice that $\mathcal{H}_3$ contains both $(i,j,k)$  and $(k,j,i)$. We have therefore proven that with probability at least $1-\delta$, when considering $M = \frac{N \left(\log\left(\frac{1}{\delta}\right) + 2\log(N)\right)}{p_{min}(1-p_{max})}$ cascades, we have $\forall (i,j,k) \in \mathcal{H}_3, \,\, \hat{h}_{i,j} > \hat{h}_{i,k} \text{ and } \hat{h}_{j,k} > \hat{h}_{i,k}.$
\end{proposition}

Putting together Proposition \ref{cl:finiteUndirected} and Corollary \ref{cor:alg}, we obtain our first theorem for learning the undirected edges with finite samples:

\begin{theorem}
With $M = \frac{N \left(\log\left(\frac{1}{\delta}\right) + 2\log(N)\right)}{p_{min}(1-p_{max})}$ cascades, with probability at least $1 -\delta$, we can learn the structure of a any bidirectional tree in the extreme-noise setting, \textit{i.e.,} when we only have access to the infection status of the nodes.
\end{theorem}

\subsection{Lower bound}
In this section, we prove a lower bound for trees in the no-noise setting. With very minor adjustments, we adapt the lower bound of \cite{Netrapalli2012}. Since for a general tree, the max degree can be up to $N-1$, we design a lower bound which is independent from the max-degree. Let $G$ be a tree drawn uniformly from $\mathcal{G}$, the set of all possible trees on $N$ nodes, and $\hat{G}$ be the reconstructed graph from the times of infection. $G \leftrightarrow T \leftrightarrow \hat{G}$ therefore forms a Markov chain. We have:
\begin{align*}
    H(G) &= I(G;\hat{G}) + H(G|\hat{G}) \\
   \text{(data processing inequality)} \quad &\leq I(G;T) + H(G|\hat{G}) \\
     \text{(independent cascades)}  \quad &\le  M\cdot I(T^1;T^{'1}) + H(G|\hat{G})\\
     \text{($I(X,Y) \leq H(X)$)}  \quad &=  M \cdot \sum_{i=1}^N H(T^1_i) + H(G|\hat{G}) \\
      \text{(Fano's inequality)} \quad &\leq  M\cdot \sum_{i=1}^N H(T^1_i) + (1 + P_e\log(|\mathcal{G}|) ) 
\end{align*}

Since $G$ is drawn uniformly from $\mathcal{G}$, $H(G) = \log(|\mathcal{G}|) $. There are $N^{N-2}$ trees on $N$ nodes, according to Cayley's formula \cite{Cayley1897}, so $H(G) = (N-2)\cdot\log(N) $.

In conclusion:
$$ M \geq \frac{(1-P_e)(N-2)\log(N)-1}{ N\cdot H(T^1_i)} = \Omega\left(\frac{\log(N)}{ H(T^1_i)}\right)$$

Using the same kind of techniques as in \cite{Netrapalli2012}, we can assume $H(T^1_i) \sim \frac{\log(N)}{N}$. Therefore:

\begin{theorem}
In the no-noise setting, we need $M = \Omega\left(N\right)$ cascades to learn the tree structure. 
\end{theorem}

In our extreme-noise setting, when we have only access to the infection status of the nodes, we can learn the tree structure with the same sample complexity (up to log-factors) as the no-noise setting!

\subsection{Tree weights} \label{sec:treeWeights}
In this section, we assume we are in the limited-noise setting, and we have access to the times of infection. We also assume we have already learned the structure of the tree.

Once we have reduced the set of possible edges by learning the structure of the bidirectional tree, learning the weights of the edges is still non-trivial. Indeed, from $T^{'m}_i$ and $T^{'m}_j$, it is still impossible to know whether this sample is useful for estimating $p_{ij}$ (case when $i$ infected $j$), or whether we should use this sample for estimating $p_{ji}$ instead (case when $j$ infected $i$). What is more, we only get one sample per node and per cascade, so it is impossible to know what really happened during that cascade. However, knowing the distribution of the noise, it is possible to compute the probability that the noise maintained the order of infections. Using this information and the reduced set of known undirected edges, we can compute two sets of $N(N-1)$ estimators, from which it is possible to infer the weights of all edges in the tree.

We introduce these two sets of $N(N-1)$ estimators, or, in other words, two estimators for each directed edge. These estimators tend to multivariate polynomials of the weights of most edges of the tree. Thus in general these polynomials have exponentially many terms; however, when $i$ and $j$ are neighbors, it is possible to express them concisely using a quantity $\mathcal{P}_{\bcancel{j}}(\rightarrow i)$, which we define formally below. This succinct representation is the key idea we exploit to solve the resulting system of equations. 

Once we know the structure of the bidirectional tree, we can consider the four estimators for each undirected edge (two estimators for each directed edge). They form a system of four equations and four unknowns, which we solve to obtain the weights of the edges.

\begin{definition}
$\mathcal{P}_{\bcancel{i}}(\rightarrow j)$ is the probability that $j$ became infected before any node on the path from $j$ to $i$, including $i$, became infected.
\end{definition}
We now introduce the estimators:
\begin{definition}
We introduce 2 sets of $N(N-1)$ estimators:
\begin{align*}
    \hat{f}_{i<j} &= \text{Fraction of infections for which  $i$ and $j$ got infected, and $i$ reported before $j$.} \\
    \hat{g}_{i,\bcancel{j}} &= \text{Fraction of infections for which $i$ got infected, but $j$ did not.}
\end{align*}
\end{definition}

By the law of large numbers, as the number of cascades scales, $\hat{f}_{i<j}$ tends to $f_{i<j}$ and $\hat{g}_{i,\bcancel{j}}$ to $g_{i,\bcancel{j}}$, where
\begin{eqnarray*}
    f_{i<j} &:=& \p(T^{'m}_i < T^{'m}_j < \infty),\\
    g_{i,\bcancel{j}} &:=& \p(T^{'m}_i < \infty, T^{'m}_j= \infty).
\end{eqnarray*}
We now compute the exact values of these two quantities. Let us assume the (unique) path between $i$ and $j$ has length $d$. We call $(u_0, \dots, u_d)$ the set of nodes on the path from $i$ to $j$, with $u_0 = i$ and $u_d = j$. We then have:
\begin{lemma}\label{lem:exactProba}
Recall $f_{i<j}$ and $g_{i,\bcancel{j}}$ are the expectation of the estimators defined above. We have:
\begin{align*}
    f_{i<j} &= \mathcal{P}_{\bcancel{j}}(\rightarrow i)\cdot p_{i\rightarrow j}\cdot s_{-d+1} + \mathcal{P}_{\bcancel{i}}(\rightarrow j)\cdot p_{j\rightarrow i}\cdot s_{d + 1} \\
    &\quad + \sum_{l=1}^{d-1}  \mathcal{P}_{\bcancel{i},\bcancel{j}}(\rightarrow u_l)\cdot p_{u_l\rightarrow i}\cdot p_{u_l\rightarrow j}\cdot s_{2l - d + 1}.\\
    \hat{g}_{i,\bcancel{j}} &= \mathcal{P}_{\bcancel{j}}(\rightarrow i)\cdot (1 - p_{i\rightarrow j}) +  \sum_{l=1}^{d-1}\mathcal{P}_{\bcancel{i},\bcancel{j}}(\rightarrow u_l)\cdot p_{u_l\rightarrow i} \cdot (1 - p_{u_l\rightarrow j}).
\end{align*}
What's more, when $i$ and $j$ are neighbors (which implies $d=1$), the expressions simplify to:
\begin{align*}
    f_{i<j} &= \mathcal{P}_{\bcancel{j}}(\rightarrow i)\cdot p_{ij}\cdot s_{0} + \mathcal{P}_{\bcancel{i}}(\rightarrow j)\cdot p_{ji}\cdot s_{2} \\
    g_{i,\bcancel{j}} &= \mathcal{P}_{\bcancel{j}}(\rightarrow i)\cdot (1 - p_{ij}).
\end{align*}
\proof \begin{align*}
    f_{i<j} &= \p(T^{'m}_i < T^{'m}_j < \infty) \\
    &= \p(\text{$i$ got infected before any nodes on the path from $i$ to $j$,} \\
    &\quad \text{then infected $j$, and the noise did not flip the times of infection}) \\
    &\quad+ \p(\text{$j$ got infected before any nodes on the path from $i$ to $j$,}\\
    &\quad  \text{then infected $i$, and the noise flipped the times of infection})\\
    &\quad+ \p(\text{One node on the path from $i$ to $j$ got infected before $i$ and $j$,} \\
    &\quad  \text{then infected both $i$ and $j$, but $i$ reported before $j$})\\
    &= \mathcal{P}_{\bcancel{j}}(\rightarrow i)\cdot p_{i\rightarrow j}\cdot s_{-d+1} \\
    &\quad+ \mathcal{P}_{\bcancel{i}}(\rightarrow j)\cdot p_{j\rightarrow i}\cdot s_{d + 1} \\
    &\quad+ \sum_{l=1}^{d-1} \mathcal{P}_{\bcancel{i},\bcancel{j}}(\rightarrow u_l)\cdot p_{u_l\rightarrow i}\cdot p_{u_l\rightarrow j}\cdot s_{2l - d + 1}.
\end{align*}
This expression is involved in general. However, if $i$ and $j$ are neighbors, then there are no nodes $u_l$ on the path between $i$ and $j$, other than $i$ and $j$ themselves. What's more, $p_{i\rightarrow j} = p_{ij}$, and $d=1$. Therefore:

$$ f_{i<j} = \mathcal{P}_{\bcancel{j}}(\rightarrow i)\cdot p_{ij}\cdot s_{0}+ \mathcal{P}_{\bcancel{i}}(\rightarrow j)\cdot p_{ji}\cdot s_{2}. $$
Let us now focus on $g_{i,\bcancel{j}}$:
\begin{align*}
    g_{i,\bcancel{j}} &= \p(T^{'m}_i < \infty, T^{'m}_j= \infty) \\
    &= \p(\text{$i$ got infected before any nodes on the path from $i$ to $j$, but $j$ did not get infected}) \\
    &\quad+ \p(\text{One node on the path from $i$ to $j$ got infected before $i$ and $j$, then infected $i$ but not $j$}) \\
    &= \mathcal{P}_{\bcancel{j}}(\rightarrow i)\cdot (1 - p_{i\rightarrow j}) \\
    &\quad+ \sum_{l=1}^{d-1}\mathcal{P}_{\bcancel{i},\bcancel{j}}(\rightarrow u_l)\cdot p_{u_l\rightarrow i} \cdot (1 - p_{u_l\rightarrow j}).
\end{align*}
As before, this expression is complex in general, but simplifies if $i$ and $j$ are neighbors, in which case:
$$ g_{i,\bcancel{j}} = \mathcal{P}_{\bcancel{j}}(\rightarrow i)\cdot (1 - p_{ij}). $$
\qed
\end{lemma}
Using the simplified expression only for when $i$ and $j$ are neighbors, we obtain:

\begin{proposition}
If we know $(i,j)$ is an edge in the original tree, then the probability of infection along this edge is given by:
$$p_{ij} = \frac{f_{i<j}\cdot s_0 - f_{j<i}\cdot s_2}{g_{i,\bcancel{j}}\cdot(s_0^2 - s_2^2) + f_{i<j}\cdot s_0 - f_{j<i}\cdot s_2}.$$
\proof 
According to Lemma \ref{lem:exactProba}, we had four second-order equations, with 4 unknowns: $p_{ij}$, $p_{ji}$, $\mathcal{P}_{\bcancel{j}}(\rightarrow i)$ and $\mathcal{P}_{\bcancel{i}}(\rightarrow j)$. We solve it, and obtain the wanted result. See Appendix \ref{app:tree} for details.
\\\qed
\end{proposition}
Combining all the pieces, we obtain our first theorem for infinite samples:

\begin{theorem}
 It is possible to learn the weights of a bidirectional tree in the limited-noise setting.
\end{theorem}

Now that we have proven the problem is solvable, we establish the number of samples needed to learn the weights with the method above.

\begin{lemma}\label{lem:treeComplexity}
With $M = \frac{N^2}{\epsilon^2}\log\left(\frac{6}{\delta}\right)\frac{\left((s_0^2 - s_2^2 + s_0 + s_2)p_{max} + s_0+s_2\right)^2}{(s_0^2 - s_2^2)^2}$ samples, with probability at least $1-\delta$, we have:
$$|\hat{p}_{ij} - p_{ij}| \leq \epsilon.$$
\proof Using Hoeffding's inequality:
\begin{align*}
    \p(|\hat{f}_{i<j} - f_{i<j}| > \epsilon_1) &\leq 2e^{-2M\epsilon_1^2},\\
    \p(|\hat{f}_{j<i} - f_{j<i}| > \epsilon_1) &\leq 2e^{-2M\epsilon_1^2},\\
    \p(|\hat{g}_{i,\bcancel{j}} - g_{i,\bcancel{j}}| > \epsilon_1) &\leq 2e^{-2M\epsilon_1^2}.
\end{align*}
Choosing $M = \frac{1}{\epsilon_1^2}\log\left(\frac{6}{\delta}\right)$, we have that with probability at least $1-\delta$, all the following hold:
\begin{align*}
    |\hat{f}_{i<j} - f_{i<j}| &\leq \epsilon_1, \\
    |\hat{f}_{j<i} - f_{j<i}| &\leq \epsilon_1, \\
    |\hat{g}_{i,\bcancel{j}} - g_{i,\bcancel{j}}| &\leq \epsilon_1.
\end{align*}
Hence, with probability at least $1-\delta$, we have (see Appendix \ref{app:tree} for details):
\begin{align*}
    \hat{p}_{ij} - p_{ij} &\leq \epsilon_1\frac{(s_0^2 - s_2^2 + s_0 + s_2)p_{max}}{g_{i,\bcancel{j}} \cdot(s_0^2 - s_2^2) + f_{i<j}\cdot s_0 - f_{j<i}\cdot s_2} \\
    &\quad + \epsilon_1 \frac{s_0+s_2}{g_{i,\bcancel{j}} \cdot(s_0^2 - s_2^2) + f_{i<j}\cdot s_0 - f_{j<i}\cdot s_2} + o(\epsilon_1 ).
\end{align*}
We use the results from Lemma \ref{lem:exactProba} to bound the denominator by $\frac{s_0^2 - s_2^2}{N}$. In the end, we obtain:
$$ |\hat{p}_{ij} - p_{ij}| \leq \epsilon_1 N \frac{(s_0^2 - s_2^2 + s_0 + s_2)p_{max} + s_0+s_2}{s_0^2 - s_2^2} + o(\epsilon_1 ).$$
We choose $\epsilon_1 =  \frac{\epsilon}{N} \frac{s_0^2 - s_2^2}{(s_0^2 - s_2^2 + s_0 + s_2)p_{max} + s_0+s_2}$. Therefore: \\
With $M = \frac{N^2}{\epsilon^2}\log\left(\frac{6}{\delta}\right)\frac{\left((s_0^2 - s_2^2 + s_0 + s_2)p_{max} + s_0+s_2\right)^2}{(s_0^2 - s_2^2)^2}$ samples, with probability at least $1-\delta$, we have $|\hat{p}_{ij} - p_{ij}| \leq \epsilon$.
\\\qed
\end{lemma}
By a union bound on all the weights of the tree, knowing there are at most $2N(N-1) < 2N^2$ directed edges in a directed tree, we obtain the following sample complexity:
\begin{theorem}
With $M = \frac{N^2}{\epsilon^2}\log\left(\frac{12N^2}{\delta}\right)\frac{\left((s_0^2 - s_2^2 + s_0 + s_2)p_{max} + s_0+s_2\right)^2}{(s_0^2 - s_2^2)^2} $ cascades, with probability $1-\delta$, we can learn all the weights of the edges of a bidirectional tree in the limited-noise setting, \textit{i.e.,} when we only have access to the noisy times of infection.
\end{theorem}

\section{Bounded-degree graphs} \label{sec:bounded}

\begin{figure} 
\centering\includegraphics[width = 0.5\linewidth]{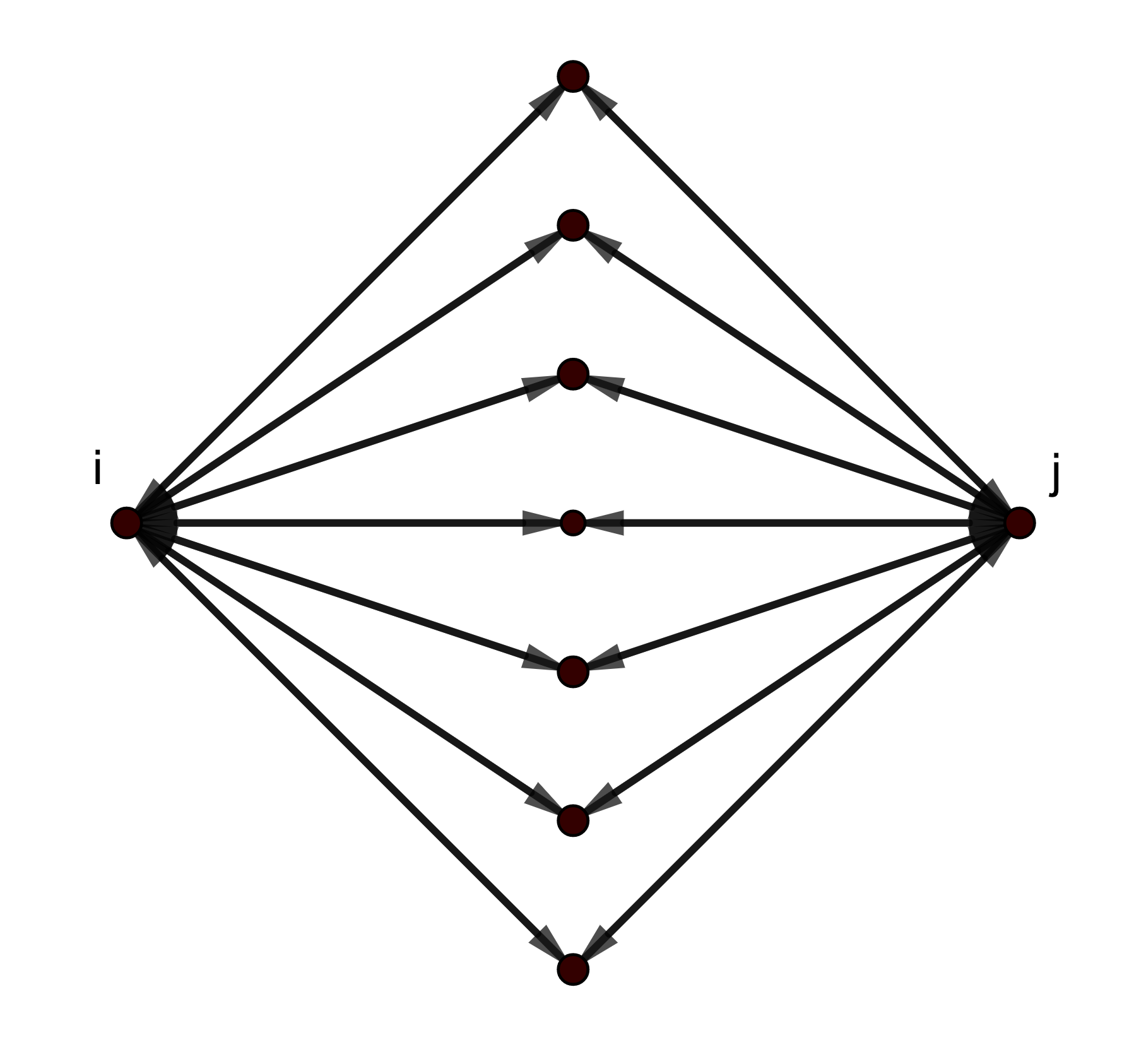}
\caption{Two nodes can be co-infected frequently without sharing an edge.}\label{fig:lotsEdges}
\end{figure}

In the previous section, the algorithm presented relies heavily on the uniqueness of paths. This property implies that we can deduce the edges from the nodes which are co-infected the most often. However, this is not true for a general bounded-degree graph. In Figure \ref{fig:lotsEdges}, we can see that the two nodes $i$ and $j$ would be co-infected frequently despite not sharing an edge. This makes the task of finding the structure much more challenging than for the bidirectional tree. 

In this section, we show  how the main ideas for learning the structure of the bidirectional tree can be extended for learning the structure of general bounded-degree graphs, in the extreme-noise setting, with almost optimal sample complexity. The framework for learning the weights of the edges in the limited-noise setting is - to the best of our knowledge - not extendable to general bounded-degree graphs; we therefore develop a new algorithm to learn the weights for general bounded-degree graphs.

\subsection{Bounded-degree structure}
In the previous section, we introduced the estimator $\hat{h}_{i,j}$, which records the fraction of cascades for which both $i$ and $j$ are infected. From a local maximum property of this estimator, we deduced the structure of the tree, in a sample efficient fashion. Indeed, if there exists a path between $i$ and $k$, and the first edge on this path is $(i,j)$, then if $i$ and $k$ are infected, $j$ must have been infected as well.  

We want to build on this idea for a bounded-degree graph of maximum degree $d$. However, for such a graph, there may be multiple paths leading from $i$ to $j$, and we cannot guarantee a single node will be infected each time. However, if $i$ is a node, $\mathcal{N}_i$ is its neighborhood, and $k \notin \mathcal{N}_i$ is another node of the graph, we can guarantee that if both $i$ and $k$ are infected, there exists a node in $\mathcal{N}_i$ which is infected. Moreover, $\mathcal{N}_i$ is the set of smallest size for which at least one node is infected at the same time as $i$ the most frequently. This leads us to a new set of estimators:

\begin{definition}
Let $i$ be a node of the graph, and let $S$ be a set such that $|S| < d$ and $i \notin S$. We define a new set of estimators:
\begin{align*}
    \hat{h}_{i,S} &= \text{Fraction of cascades for which $i$ is infected, and at least one node of S is infected.}
\end{align*}
\end{definition}
Let us assume that for each pair $(i,S)$, we have access to the limit $h_{i,S}$ of $\hat{h}_{i,S}$. We now introduce an algorithm showing how to leverage these limits to learn the structure of any bounded-degree graph.
\begin{algorithm}[H] 
\caption{Learn the undirected edges of any graph of maximum degree $d$.}\label{algo:generalGraph}
\begin{algorithmic}[1]
\Procedure{LearnGraph}{$\{h_{i,S}\}_{i \in V}^{|S| \leq d}$} 
\State $edges \gets []$
\For{$i=1\dots n_{nodes}$}
\State $S\_max\_i \gets$ set such that $h_{i, S\_max\_i}$ is maximal, and such that $\textsc{Size}(S\_max\_i)$ is minimal. 
\For{$n_j$ in $S\_max\_i$}
\State Add edge $(i, n_j)$ to $edges$
\EndFor
\EndFor
\Return $edges$
\EndProcedure
\end{algorithmic}
\end{algorithm}

We show this algorithm is correct.

\begin{lemma}
    Algorithm \ref{algo:generalGraph} correctly finds the neighborhood of each node.
    \proof Let us recall that $h_{i,S}$ is the probability that node $i$ and at least one node of $S$ are co-infected. To prove the correctness of this algorithm, it suffices to prove:
    
    $$ \forall i \in V, \forall S \,\, s.t. \, |S|\leq d, \;\; h_{i, S} \geq  h_{i, \mathcal{N}_i} \implies \mathcal{N}_i \subseteq S$$ 
Let pick a set $S$ such that $ \mathcal{N}_i \setminus S \neq \emptyset$, and let $k$ be a node in  $\mathcal{N}_i \setminus S$. We know $h_{i,S\cup \{k\}} \geq h_{i,S} + \p(\text{$i$ and $k$ are the only infected nodes})$. Since $i$ and $k$ are neighbors, $\p(\text{$i$ and $k$ are the only infected nodes}) > 0$, and therefore $h_{i,S\cup \{k\}} > h_{i,S}$. Following this line of reasoning, if  $ \mathcal{N}_i \setminus S \neq \emptyset$, $S$ we can always increase the value of $h_{i,S}$ by adding a node of $\mathcal{N}_i$.

However, it is impossible to increase the value of $h_{i, \mathcal{N}_i}$, because if $i$ and any other node of the graph are co-infected, we know one node of $\mathcal{N}_i$ is also infected. Therefore, the algorithm is correct.
\\\qed
\end{lemma}

Unfortunately, we do not have direct access to $h_{i,S}$. We therefore study how many samples are needed to replace $h_{i,S}$ by its estimate $\hat{h}_{i,S}$ while preserving the correctness of the algorithm. Just like in Section \ref{sec:treeStruct}, we notice that we do not need the $\{\hat{h}_{i,S}\}_{i \in V}^{|S| \leq d}$ to be close to their limit, we only need the indexes of their ordering to be the same. We notice that the neighborhood $\mathcal{N}_i$ of $i$ is the set of smallest size which is the most often infected at the same time as $i$ is ($\mathcal{N}_i = \displaystyle\argmax_{|R| \leq d} h_{i,R}$). Therefore, we must have that for every set $S$, $\hat{h}_{i,S} \leq \hat{h}_{i,\mathcal{N}_i}$. However, some other sets might achieve the same value if we do not observe enough cascades. This could happen in two cases:
\begin{itemize}
    \item Not all the nodes of the neighborhood were infected, and therefore a subset $T_1 \subset \mathcal{N}_i$ of the neighborhood is such that $\hat{h}_{i,T_1} = \hat{h}_{i,\mathcal{N}_i}$. Since $|T_1| < |\mathcal{N}_i|$, the algorithm would return $T_1$ and not $\mathcal{N}_i$, which would be a failure case.
    \item Some other node $k$ is always infected every time a specific  node $j \in \mathcal{N}_i$ is infected. The set $T_2 = \mathcal{N}_i \setminus \{j\} \cup \{k\}$ will therefore be such that $\hat{h}_{i,T_2} = \hat{h}_{i,\mathcal{N}_i}$, and the algorithm would not know which maximum to pick. This is a failure case as well.
\end{itemize}
 We identify events which guarantee that the failure cases above cannot arise. Let $E_{i,j}^m$ be the event that $i$ and $j$ were the only infected nodes during cascade $m$. $E_{i,j} = \displaystyle\bigcup_{1 \leq m \leq M} E_{i,j}^m$ is therefore the event that there exists a cascade for which only $i$ and $j$ were infected. If such a cascade exists, the set $S_{max} = \displaystyle\argmax_{|R| \leq d} \hat{h}_{i,R}$ \textit{must} contain $j$. If the event $E_{i,j}$ happens for every node $j$ in the neighborhood of $i$, we can therefore guarantee Algorithm \ref{algo:generalGraph} is correct. We characterize the sample complexity needed for this below.

\begin{proposition} \label{cl:oneCascade}
Let $i$ be a node, and let $j$ be one of its neighbors. Let $S$ be a set of size $|S| \leq d$, such that $i \notin S$ and $j \notin S$. With probability at least $1-\frac{\delta}{d\cdot N^{d+2}}$, among $M= \frac{(d+2)\cdot N\log(N) + N\log \left(\frac{d}{\delta}\right)}{p_{min}(1 - p_{max})^{2(d-1)}}$ cascades, there exists a cascade in which $i$ and $j$ are infected, but no nodes of $S$ are infected.
\proof We notice that if, during cascade $m$, the only infected nodes are $i$ and $j$, no nodes of $S$ are infected. This is exactly the event $E_{i,j}^m$ defined above.
\begin{align*}
    \p(E_{i,j}^m) 
     &= \p(\text{$i$ and $j$ are the only infected nodes during cascade $m$}) \\
     &\geq \p(\text{cascade started at $i$, $j$ was the only infected neighbor of $i$, and $j$ did not infect any nodes}) \\
    &\geq \frac{1}{N} \cdot p_{min} \cdot (1-p_{max})^{2(d-1)}.
\end{align*}
The probability that this never happens during $M$ cascade is exactly the complement of the event $E_{i,j}$ defined above:
\begin{align*}
    \p(\textsc{Not}(E_{i,j})) 
    &\leq \left(1- \frac{1}{N} \cdot p_{min} \cdot (1-p_{max})^{2(d-1)} \right)^M \\
    &\leq e^{-M\cdot \frac{p_{min}}{N} \cdot (1-p_{max})^{2(d-1)}} \\
    &\leq \frac{\delta}{d\cdot N^{d+2}}.
\end{align*}
\qed
\end{proposition}

\begin{lemma}
    With probability at least $1-\delta$, if we observe $M= \frac{(d+2)\cdot N\log(N) + N\log \left(\frac{d}{\delta}\right)}{p_{min}(1 - p_{max})^{2(d-1)}}$ cascades, Algorithm \ref{algo:generalGraph} is correct, and has running time $\OO(M \cdot N^{d+2}) = \OO(d \cdot N^{d+3}\log(N))$.
    
    \proof Let $\mathcal{S} = \{ S \in \mathcal{P}(V), |S| \leq d \}$ be the set of sets of nodes of size at most $d$,  let $\mathcal{S}_{\bcancel{i}} = \{S \in \mathcal{S}, i \notin S \}$ be the set of sets of $\mathcal{S}$ which do not contain $i$, and let $\mathcal{N}_i = \{ j, (i,j) \in E \}$ be the neighborhood of $i$. We pick a set $S  \in \mathcal{S}$, such that $S \neq \mathcal{N}_i$. 
    
    We first prove that $\hat{h}_{i, S} \leq \hat{h}_{i, \mathcal{N}_i}$. Since the neighborhood of $i$ separates $i$ from the rest of the graph, and infected nodes are connected, we can conclude that every time $i$ and another node of $S$ are infected, one node in the neighborhood $\mathcal{N}_i$ of $i$ is infected as well. Therefore, we cannot increase $\hat{h}_{i, S}$ without also increasing $\hat{h}_{i, \mathcal{N}_i}$. In particular, this means that even if $|S| > |\mathcal{N}_i|$ (for instance if $\mathcal{N}_i \subset S$), we have $\hat{h}_{i, S} \leq \hat{h}_{i, \mathcal{N}_i}$. We therefore know that $ \mathcal{N}_i \in \displaystyle\argmax_{R \in \mathcal{S}_{\bcancel{i}} } \hat{h}_{i, R} $.
    
    We now prove that with probability at least $1-\delta$, $\mathcal{N}_i$ is the set of $\displaystyle\argmax_{R \in \mathcal{S}_{\bcancel{i}} } \hat{h}_{i, R} $ of minimal size. To do so, we notice that if there exists a cascade such that $i$ and $j \in \mathcal{N}_i$ are infected, but no node of $S$ is infected, then this implies $\hat{h}_{i, S} < \hat{h}_{i, \mathcal{N}_i}$. Indeed, as shown above, every time we increase $\hat{h}_{i, S}$, we also increase $\hat{h}_{i, \mathcal{N}_i}$, and we know there exists one cascade for which we increased $\hat{h}_{i, \mathcal{N}_i}$ without increasing $\hat{h}_{i, S}$. We now calculate the probability $P_{\rm failure}$ that such a cascade does not exist for all nodes $i$ in the graph, all nodes $j$ in their neighborhood, and all sets $S$ which do not include $i$ or $j$. 
    \begin{align*}
       P_{\rm failure} &\leq \p(\exists i \in V, \exists j \in \mathcal{N}_i, \exists S \in \mathcal{S}_{\bcancel{i}, \bcancel{j}}, \text{every time $i$ and $j$ are infected, a node of $S$ is also infected}) \\ 
       &\leq \sum_{i\in V} \sum_{j \in \mathcal{N}_i} \sum_{S \in \mathcal{S}_{\bcancel{i}, \bcancel{j}}} \p(\text{every time $i$ and $j$ are infected, a node of $S$ is also infected}).
    \end{align*}
    We use Proposition \ref{cl:oneCascade} to bound this quantity:
    \begin{align*}
       P_{\rm failure} 
       &\leq \sum_{i\in V} \sum_{j \in \mathcal{N}_i} \sum_{S \in \mathcal{S}_{\bcancel{i}, \bcancel{j}}} \frac{\delta}{d\cdot N^{d+2}} \\
       &\leq N \cdot d \cdot N^{d+1} \cdot \frac{\delta}{d\cdot N^{d+2}} \\
       &\leq \delta.
    \end{align*}
We now know that with probability at least $1-\delta$: $$\forall R \in \bigcup_{j \in \mathcal{N}_i} \mathcal{S}_{\bcancel{i}, \bcancel{j}},\quad \hat{h}_{i, R} < \hat{h}_{i, \mathcal{N}_i}.$$
This implies that no set of $\bigcup_{j \in \mathcal{N}_i} \mathcal{S}_{\bcancel{i}, \bcancel{j}}$ can belong in $\displaystyle\argmax_{R \in \mathcal{S}_{\bcancel{i}} } \hat{h}_{i, R}$. However, we have:
    $$ \mathcal{S}_{\bcancel{i}} \setminus \bigcup_{j \in \mathcal{N}_i} \mathcal{S}_{\bcancel{i}, \bcancel{j}} =  \{S \in \mathcal{S}, \mathcal{N}_i \subseteq S \}. $$
    In particular, this means that $\mathcal{N}_i$ is the only set of $ \mathcal{S}_{\bcancel{i}} \setminus \bigcup_{j \in \mathcal{N}_i} \mathcal{S}_{\bcancel{i}, \bcancel{j}}$ of minimal size. This shows that with probability at least $1-\delta$, $\mathcal{N}_i$ is the set of $\displaystyle\argmax_{R \in \mathcal{S}_{\bcancel{i}} } \hat{h}_{i, R} $ of minimal size. This proves Algorithm \ref{algo:generalGraph} is correct, and that we can learn the structure of any graph of maximum degree $d$ with $M= \frac{(d+2)\cdot N\log(N) + N\log \left(\frac{d}{\delta}\right)}{p_{min}(1 - p_{max})^{2(d-1)}}$ cascades.
    
    Since we do at most one operation by pair of (node, set) and by cascade, the running time is $\OO(N\cdot N^{d+1} \cdot M)$, which is what we wanted to prove.
    \\\qed
\end{lemma}

This leads to our theorem for learning the structure of any bounded-degree graph.

\begin{theorem}
    With probability at least $1-\delta$, in the extreme-noise setting, we can learn the structure of any graph of maximum degree $d$ with $M= \OO\left(\frac{d\cdot N\log\left(\frac{N}{\delta}\right) }{p_{min}(1 - p_{max})^{2d}}\right)$ cascades in polynomial time.
\end{theorem}    

Let us now assume that $p_{max} \sim \frac{1}{d}$. This assumption is reasonable when you expect a constant number of infections by time step. For instance, it makes sense for real diseases, for which carriers have to meet to transmit it (we can only meet a constant number of people each day). It would not make sense for social networks, in which it is possible to reach many followers with each post. 

\begin{corollary}
    With probability at least $1-\delta$, in the extreme-noise setting, if we assume $p_{max} \sim \frac{1}{d}$ and $p_{min}$ constant, we can learn the structure of any graph of maximum degree $d$ with $M= \OO\left(d\cdot N\log\left(\frac{N}{\delta}\right) \right)$. This sample complexity is optimal up to log-factors, and almost matches the lower bound established in the no-noise setting.
    \proof We need at least $\OO\left(d\cdot N\log\left(N\right)\right)$ samples to learn the structure of a bounded-degree graph with maximum degree $d$, according to the lower bound in \cite{Netrapalli2012}. 
\end{corollary}

\subsection{Bounded-degree weights} \label{sec:cascade2}
For the remainder of this paper, we state results in the limited-noise setting.

If we consider cascades of size $k$, the exact probability of infection between two nodes is a multivariate polynomial of degree $N$ on $N(N-1)$ variables (the variables here would be the weights of the graph), with a sum of up to $2\cdot\displaystyle \sum_{l = 1}^{k} \frac{(N-2)!}{(N-k)!}$ terms. If the graph has more than five nodes, the resulting polynomial is of degree more than five.

For our algorithm, we therefore only use cascades of size 1 or 2. This is a waste of the data, since we simply discard cascades of larger size. However, we are not aware of techniques on how to utilize larger cascades in the limited-noise setting. Cascades of size 1 or 2 are simple enough that we can write explicitly their probability. This allows use to:
\begin{enumerate}
    \item Design estimators for which we can calculate the exact limit.
    \item Combine these estimators to transform a polynomial system of degree $N$ to a polynomial system of degree 2.
    \item Solve this system exactly and obtain the probabilities of infection.
\end{enumerate}

\subsection{Estimators}
We start by designing a few estimators:

\begin{definition}
We introduce two sets of $N(N-1)$ estimators and one set of $N$ estimators. These estimators can be computed even if we only have access to the noisy times of infection.
\begin{align*}
    \hat{h}^2_{i,j} &= \text{Fraction of cascades for which only $i$ and $j$ are infected.} \\
    \hat{f}^2_{i<j} &= \text{Fraction of cascades for which only $i$ and $j$ are infected, and $t'_i < t'_j$.} \\
    \hat{e}^1_i &= \text{Fraction of cascades for which only $i$ is infected.} 
\end{align*}
\end{definition}
To simplify the coming notations, let us introduce $s_k$, which is the probability that the noise on $j$ has delay at least $k$ relative to the noise on $i$. Since the noise is i.i.d., this value is independent from $i$ and $j$: $s_k = \p(n_j - n_i \geq k)$\footnote{For instance, for geometric noise of parameter $q$, we have: 
$s_k = \displaystyle\sum_{t_j = \max(0, k)}^{\infty}\sum_{t_i = 0}^{t_j - k} (1-q)^{t_i + t_j} = (1-q)^{\max(0, k)}\left(1 - \frac{(1-q)^{1 - \min(0,k)}}{2-q} \right).$}. For instance, if $i$ infected $j$ during cascade $m$, the probability that the noise did not flip the order of infection (\textit{i.e.} $T^{'m}_i < T^{m}_j$) is $\p(n_j \geq n_i) = s_0$. In the reverse case, the probability that the noise flipped the order of infection is $\p(T^{m}_j +  n_j < T^{'m}_i +  n_i) = \p(1 +  n_j <  n_i) = \p( n_i -  n_j \geq 2) = s_2$.

We now compute the limits of those estimators. 

\begin{proposition} \label{prop:estimatorsCascade2}
As the number of cascades $M$ goes to infinity, the estimators introduced above tend to the following limit:
\begin{align*}
    \hat{h}^2_{i,j} &\to_{M\to\infty} \frac{1}{N}(p_{ij} + p_{ji})\prod_{k \neq i,j} (1-p_{ik})(1-p_{jk}), \\
    \hat{f}^2_{i<j} &\to_{M\to\infty} \frac{1}{N}(p_{ij}\cdot s_0 + p_{ji}\cdot s_2)\prod_{k \neq i,j} (1-p_{ik})(1-p_{jk}), \\
    \hat{e}^1_i &\to_{M\to\infty} \frac{1}{N}(1-p_{ij})\prod_{k \neq i,j} (1-p_{ik}).
\end{align*}
\proof The proof is very similar to \ref{lem:exactProba}. See details in Appendix \ref{sec:appBoundedComplexity}.\\\qed
\end{proposition}

\subsection{Solving the system}
The limit of these estimators, as seen as a function of the probabilities of infection, is a complex polynomial on up to $2(N-1)$ variables. The crux of our algorithm is to combine those estimators in order to cancel out most of these variables, and create $N-1$ systems of two equations of degree 2 and two unknowns, which we then solve.

\begin{proposition}
Let $\hat{V}_{ij} = \frac{\hat{f}^2_{i<j}}{\hat{h}^2_{i,j} + N\cdot \hat{e}^1_i \cdot \hat{e}^1_j}$. Then the limit of $\hat{V}_{ij}$ as the number of cascades $M$ goes to infinity only depends on the variables $p_{ij}$ and $p_{ji}$ :
$$\hat{V}_{ij} \to_{M\to\infty} \frac{p_{ij}\cdot s_0 + p_{ji}\cdot s_2}{1 + p_{ij}\cdot p_{ji}}. $$
\proof Since all the estimators converge towards a constant, we can use Slutsky's lemma to find the limit of $\hat{V}_{ij}$. Let $V_{ij}$ be the limit of $\hat{V}_{ij}$ as $M$ goes to infinity. We notice that the $\displaystyle\prod_{k \neq i,j} (1-p_{ik})(1-p_{jk})$ parts cancel each other out:
\begin{align*}
    V_{ij} &= \frac{f^2_{i<j}}{h^2_{i,j} + N\cdot e^1_i \cdot e^1_j} \\
    &=\frac{\frac{1}{N}(p_{ij}\cdot s_0 + p_{ji}\cdot s_2)\left[\displaystyle\prod_{k \neq i,j} (1-p_{ik})(1-p_{jk})\right]}{\left(\frac{1}{N}(p_{ij} + p_{ji}) + N \frac{(1-p_{ij})(1-p_{ji})}{N^2}\right)\left[\displaystyle\prod_{k \neq i,j} (1-p_{jk})(1-p_{jk})\right]} \\
    &= \frac{(p_{ij}\cdot s_0 + p_{ji}\cdot s_2)}{1+ p_{ij}\cdot p_{ji}}.
\end{align*}
\qed
\end{proposition}

We can therefore use this equality to deduce the weights of all the edges of the graph:

\begin{theorem}
For any graph, for any noise distribution having finite values, we can learn the weights of all the edges of the graph. In particular, we can compute a quantity which converges to the true weight of each edge:
$$ \hat{p}_{ij} = \frac{2(\hat{V}_{ji}s_2 - \hat{V}_{ij}s_0)}{\left(s_0^2 - s_2^2\right) + \sqrt{\left(s_0^2 - s_2^2\right)^2 - 4(\hat{V}_{ji}s_2 - \hat{V}_{ij}s_0)(\hat{V}_{ij}s_2-\hat{V}_{ji}s_0)}}.$$
\proof We present a sketch of the proof here. The details can be found in Appendix \ref{app:bounded}. We know $\hat{V}_{ij}$ tends to $V_{ij} = \frac{p_{ij}\cdot s_0 + p_{ji}\cdot s_2}{1+ p_{ij}\cdot p_{ji}}$. Using both $V_{ij}$ and $V_{ji}$, we can establish this second-degree equation:
$$V_{ji}s_2 - V_{ij}s_0 + \left(s_0^2 - s_2^2\right) p_{ij} +\left(V_{ij}s_2-V_{ji}s_0 \right)p_{ij}^2 = 0  $$
We recall that by definition, $s_0 \geq s_2$. We also notice that if $p_{ij} = q_1$ and $p_{ji} = q_2$ is a pair of solutions of this system, then $p_{ij} = \frac{1}{q_2}$ and $p_{ji} = \frac{1}{q_1}$ forms the other pair of solution, which implies there is uniqueness of solutions in $[0,p_{max}]$. Since the real probabilities of infection satisfy this system, we also know the solution exists. Let $\Delta = \left(s_0^2 - s_2^2\right)^2 - 4(V_{ji}s_2 - V_{ij}s_0)(V_{ij}s_2-V_{ji}s_0)$. The only solution of this system in $[0,p_{max}]$ is then:
\begin{align*}
    p_{ij} &=\frac{2(V_{ji}s_2 - V_{ij}s_0)}{\left(s_0^2 - s_2^2\right) + \sqrt{\Delta}}.
\end{align*}
\qed
\end{theorem}

\subsection{Sample complexity}
We establish the sample complexity needed to estimate $p_{ij}$ with precision $\epsilon$. To do so, we start by estimating $V_{ij}$. Note that we only consider the pair of nodes $i$ and $j$ if, among the $M$ samples, there exists a cascade of size 2 in which $i$ and $j$ are the only infected nodes (\textit{i.e.} $h_{i,j}^2 > 0$). Otherwise, we set $\hat{p}_{ij} = \hat{p}_{ji} = 0$ as our estimate for $p_{ij}$.

\begin{proposition} \label{prop:epsilonV}
With probability $1- \frac{\delta}{N^2}$, with $M = $ samples, we can estimate $V_{ij}$ with precision $\epsilon_V$.
\proof We present a sketch of the proof; the details can be found in Appendix \ref{sec:appBoundedComplexity}. As in Proposition \ref{lem:treeComplexity}, we use Hoeffding's inequality:
\begin{align*}
    \p(|\hat{f}^2_{i<j} - f^2_{i<j}| >\epsilon_1) &\leq 2e^{-2M\epsilon_1^2},\\
    \p(|\hat{h}^2_{i,j} - h^2_{i,j}| >\epsilon_1) &\leq 2e^{-2M\epsilon_1^2},\\
    \p(|\hat{e}^1_{i} - e^1_{i}| >\epsilon_1) &\leq 2e^{-2M\epsilon_1^2}, \\
    \p(|\hat{e}^1_{j} - e^1_{j}| >\epsilon_1) &\leq 2e^{-2M\epsilon_1^2}.
\end{align*}
We use this to bound above $V_{ij}$:
\begin{align*}
\hat{V}_{ij} &\leq V_{ij} \left[ 1 + \frac{\epsilon_1}{f^2_{i<j}} +\epsilon_1\frac{1 + N(e^1_i + e^1_j)}{h^2_{i,j} + Ne^1_i e^1_j} + o(\epsilon_1) \right]
\end{align*}
By bounding below the denominators and bounding above the numerator, we finally obtain:
\begin{align*}
|\hat{V}_{ij} - V_{ij}| 
&\leq \epsilon_1\frac{4N}{p_{min}s_2(1-p_{max})^{2d}} + o(\epsilon_1).
\end{align*}
Therefore, by union bound, and by choosing  $\epsilon_1 \frac{4N}{p_{min}s_2(1-p_{max})^{2d}}= \epsilon_V$, and setting $2 e^{-2M\epsilon_1^2} = \frac{\delta}{3N^2}$, we obtain: \\
With $M =  \frac{1}{\epsilon_V^2} \frac{16N^2}{p_{min}^2s_2^2(1-p_{max})^{4d}}\frac{2\log(3N) - \log(\delta)}{2}$ samples, we can guarantee $|\hat{V}_{ij} - V_{ij}| \leq \epsilon_V $ with probability at least $1 - \frac{\delta}{N^2}$.
\\\qed
\end{proposition}

Once we have estimated $V_{ij}$ with precision $\epsilon_V$, estimating $p_{ij}$ is unfortunately still not an easy task. Indeed, let $\Delta = \left(s_0^2 - s_2^2\right)^2 - 4(V_{ji}s_2 - V_{ij}s_0)(V_{ij}s_2-V_{ji}s_0)$. We have $p_{ij} = \frac{-\left(s_0^2 - s_2^2\right) + \sqrt{\Delta}}{2(V_{ij}s_2-V_{ji}s_0)} $, which means that $\Delta$ has to be positive for this quantity to be defined. However, for general values of $V_{ij}$ and $V_{ji}$, $\Delta$ can be negative. We therefore use the framework of constrained optimization to bound $\Delta$ away from 0.

\begin{lemma}\label{lem:delta}
Let $\Delta = \left(s_0^2 - s_2^2\right)^2 - 4(V_{ji}s_2 - V_{ij}s_0)(V_{ij}s_2-V_{ji}s_0)$. We have:
$$\Delta \geq (s_0^2 - s_2^2)^2\frac{(1 - p_{max})^2}{1 + p_{max}^2}.$$
\proof We find the lower bound on $\Delta$ by reformulating the problem as a constrained optimization problem, and introducing the corresponding Lagrangian multipliers. The details can be found in Appendix \ref{app:bounded}.
\\\qed
\end{lemma}

Now that we have established this bound on $\Delta$, we can give the sample complexity needed to estimate $p_{ij}$.

\begin{proposition} \label{prop:epsilon}
Assuming we can estimate $V_{ij}$ within precision $\epsilon_V$, then we can estimate $p_{ij}$ within precision $\epsilon = \frac{6\epsilon_V(1 + p_{max}^2)}{(s_0^2 - s_2^2)^2(1 - p_{max})^2}$.
\proof This is a simple derivation which can be found in Appendix \ref{sec:appBoundedComplexity}. \\\qed
\end{proposition}

We finally piece everything together, and use a union bound on all the $p_{ij}$ to obtain the final sample complexity of our algorithm for learning the weights of general bounded-degree graphs.

\begin{theorem}
In the limited-noise setting, with probability at least $1-\delta$, with $M =  \OO\left( \frac{e^{4p_{max}(d+1)}}{p_{min}^2s_2^2(s_0^2 - s_2^2)^4} \frac{N^2}{\epsilon^2} \log\left(\frac{N}{\delta}\right) \right)$ samples, we can learn the weights of any bounded-degree graph up to precision $epsilon$.
\proof We obtain the desired bound by combining Proposition \ref{prop:epsilonV}
and Proposition \ref{prop:epsilon}. See details in Appendix \ref{sec:appBoundedComplexity}.\\\qed
\end{theorem}

Once again, if we assume $p_{max} \sim \frac{1}{d}$ and $p_{min}$ constant, we obtain the following sample complexity:

\begin{corollary}
In the limited-noise setting, with probability at least $1-\delta$, if $p_{max} \sim \frac{1}{d}$ and $p_{min}$ constant, we can learn the weights of any bounded-degree graph up to precision $epsilon$ with $M =  \OO\left( \frac{1}{s_2^2(s_0^2 - s_2^2)^4} \frac{N^2}{\epsilon^2} \log\left(\frac{N}{\delta}\right) \right)$ samples.
\end{corollary}

\section{General graphs}
In the limited-noise setting, we notice that nothing prevents us from using the algorithm for learning bounded-degree weights for general graph. This proves this problem is solvable for $any$ graph, and $any$ noise distribution. As explained in Section \ref{sec:hard}, this is not an obvious result. However, if we do not assume $p_{max} \sim \frac{1}{N}$, the sample complexity is now exponential.

\begin{theorem}
In the limited-noise setting, with probability at least $1-\delta$, it is possible to learn all the weights of $any$ graph, for $any$ noise distribution, with finite (but potentially exponential) sample complexity.
\end{theorem}

\section{Discussion and Future work}
In this paper, we presented the first results to learn the edges of a graph from noisy times of infection. We showed we learn the structure of any bidirectional tree or any bounded-degree graph (note that not all trees are bounded-degree graphs) with optimal sample complexity (up to log-factors).

However, our results are not tight for learning the weights of bidirectional trees or bounded-degree graphs. In the no-noise setting, \cite{Netrapalli2012} proves this can be achieved with $\OO(d^2N\log(N))$ for graphs of maximum degree $d$ with correlation decay. Our results have sample complexity $\OO(N^2\log(N))$ without any assumption of correlation decay. It is an open problem to understand whether it is possible to achieve a sample complexity of $\OO(d^2N\log(N))$ in the limited-noise setting, and whether assuming correlation decay is necessary to obtain such a result.

Moreover, for learning the weights of a general bounded-degree graph, we only use cascades of size 1 or 2. If we are given an infinite number of cascades of size bigger than 2, our current algorithm cannot learn the weights of the graph. Future work could develop an algorithm without such a weakness.

All our results for learning the weights of the edges are in the limited-noise setting. Whether or not it is possible to learn the noise in the extreme-noise setting is an other question of interest for future work.

Finally, we have made no restriction on the distribution of the noise we add, other than it is finite. It would be interesting to study whether stronger restrictions on the noise (for instance Gaussian noise) would lead to stronger results. It would also be interesting to allow infinite noise, and develop algorithms which are robust to errors in the infection status of a node (our current algorithms can return wrong graph structure with only one adversarially chosen false positive).

\newpage
\bibliography{all}

\begin{thebibliography}{10}

\bibitem{Abrahao2013}
Bruno Abrahao, Flavio Chierichetti, Robert Kleinberg, and Alessandro Panconesi.
\newblock {Trace complexity of network inference}.
\newblock {\em Proceedings of the 19th ACM SIGKDD international conference on
  Knowledge discovery and data mining - KDD '13}, page 491, 2013.

\bibitem{Arias-castro2011}
Ery Arias-castro, Emmanuel~J Cand{\`{e}}s, and Arnaud Durand.
\newblock {Detection of an anomalous cluster in a network}.
\newblock {\em The Annals of Statistics}, 39(1):278--304, 2011.

\bibitem{Arias-castro}
Ery Arias-castro and S~T Nov.
\newblock {Detecting a Path of Correlations in a Network}.
\newblock pages 1--12.

\bibitem{Bernoulli2004}
Daniel Bernoulli and Sally Blower.
\newblock {An attempt at a new analysis of the mortality caused by smallpox and
  of the advantages of inoculation to prevent it}.
\newblock {\em Reviews in medical virology}, 14:275--288, 2004.

\bibitem{Cayley1897}
A.~Cayley.
\newblock {A theorem on trees}.
\newblock In {\em Collected Mathematical Papers Vol. 13}, pages 26--28.
  Cambridge University Press, 1897.

\bibitem{Cheng2014}
Justin Cheng, Lada~A. Adamic, P.~Alex Dow, Jon Kleinberg, and Jure Leskovec.
\newblock {Can Cascades be Predicted?}
\newblock In {\em Proceedings of the 23rd international conference on World
  wide web (WWW' 14)}, 2014.

\bibitem{DelVicario2016}
Michela {Del Vicario}, Alessandro Bessi, Fabiana Zollo, Fabio Petroni, Antonio
  Scala, Guido Caldarelli, H.~Eugene Stanley, and Walter Quattrociocchi.
\newblock {The spreading of misinformation online}.
\newblock {\em Proceedings of the National Academy of Sciences}, page
  201517441, 2016.

\bibitem{Dempster1977}
A.~P. Dempster, N.~M. Laird, and D.~B. Rubin.
\newblock {Maximum Likelihood from Incomplete Data via the EM Algorithm}.
\newblock {\em Journal ofthe Royal Statistical Society}, 39(1):1--38, 1977.

\bibitem{Drakopoulos2014}
Kimon Drakopoulos, Asuman Ozdaglar, and John~N. Tsitsiklis.
\newblock {An efficient curing policy for epidemics on graphs}.
\newblock {\em arXiv preprint arXiv:1407.2241}, (December):1--10, 2014.

\bibitem{Drakopoulos2015}
Kimon Drakopoulos, Asuman Ozdaglar, and John~N. Tsitsiklis.
\newblock {A lower bound on the performance of dynamic curing policies for
  epidemics on graphs}.
\newblock (978):3560--3567, 2015.

\bibitem{fanti2016rumor}
Giulia Fanti, Peter Kairouz, Sewoong Oh, Kannan Ramchandran, and Pramod
  Viswanath.
\newblock {Rumor source obfuscation on irregular trees}.
\newblock In {\em Proceedings of the 2016 ACM SIGMETRICS International
  Conference on Measurement and Modeling of Computer Science (SIGMETRICS' 16
  )}, pages 153--164. ACM, 2016.

\bibitem{Fanti2017}
Giulia Fanti, Peter Kairouz, Sewoong Oh, Kannan Ramchandran, and Pramod
  Viswanath.
\newblock {Hiding the Rumor Source}.
\newblock {\em IEEE Transactions on Information Theory}, 63(10):6679--6713,
  2017.

\bibitem{Fanti2014}
Giulia Fanti, Peter Kairouz, Sewoong Oh, and Pramod Viswanath.
\newblock {Spy vs. Spy: Rumor Source Obfuscation}.
\newblock {\em Proceedings of the 2015 ACM SIGMETRICS International Conference
  on Measurement and Modeling of Computer Systems (SIGMETRICS' 14)}, pages
  271--284, 2015.

\bibitem{Farajtabar2017}
Mehrdad Farajtabar, Jiachen Yang, Xiaojing Ye, Huan Xu, Rakshit Trivedi, Elias
  Khalil, Shuang Li, Le~Song, and Hongyuan Zha.
\newblock {Fake News Mitigation via Point Process Based Intervention}.
\newblock In {\em Proceedings of the 34th International Conference on Machine
  Learning (ICML' 17)}, 2017.

\bibitem{Goldberg2001}
Ken Goldberg, Theresa Roeder, Dhruv Gupta, and Chris Perkins.
\newblock {Eigentaste: A Constant Time Collaborative Filtering Algorithm}.
\newblock {\em Information Retrieval}, 4(2):133--151, 2001.

\bibitem{Gomez-rodriguez2012}
Manuel Gomez-rodriguez, Jure Leskovec, and Andreas Krause.
\newblock {Inferring Networks of Diffusion and Influence}.
\newblock In {\em ACM Transactions on Knowledge Discovery from Data (TKDD'
  12)}, volume~5, 2012.

\bibitem{Gomez-Rodriguez2013}
Manuel Gomez-Rodriguez, Jure Leskovec, and Bernhard Sch{\"{o}}lkopf.
\newblock {Structure and Dynamics of Information Pathways in Online Media}.
\newblock In {\em 6th International Conference on Web Search and Data Mining
  (WSDM 2013)}, 2013.

\bibitem{Hoffmann2018}
Jessica Hoffmann and Constantine Caramanis.
\newblock {The Cost of Uncertainty in Curing Epidemics}.
\newblock {\em Proceedings of the ACM on Measurement and Analysis of Computing
  Systems (SIGMETRICS' 18)}, 2(2):11--13, 2018.

\bibitem{Iwata2013}
Tomoharu Iwata, Amar Shah, and Zoubin Ghahramani.
\newblock {Discovering Latent Influence in Online Social Activities via Shared
  Cascade Poisson Processes}.
\newblock In {\em Proceedings of the 19th ACM SIGKDD international conference
  on Knowledge discovery and data mining (KDD' 13)}, 2013.

\bibitem{Kempe2003}
David Kempe, Jon Kleinberg, and {\'{E}}va Tardos.
\newblock {Maximizing the spread of influence through a social network}.
\newblock In {\em Proceedings of the ninth ACM SIGKDD international conference
  on Knowledge discovery and data mining - KDD '03}, 2003.

\bibitem{Khim2017}
Justin Khim and Po-Ling Loh.
\newblock {Permutation Tests for Infection Graphs}.
\newblock pages 1--28, 2017.

\bibitem{Khim2018}
Justin Khim and Po-Ling Loh.
\newblock {A theory of maximum likelihood for weighted infection graphs}.
\newblock pages 1--47, 2018.

\bibitem{Kwon2019}
Jeongyeol Kwon, Wei Qian, Constantine Caramanis, Yudong Chen, and Damek Davis.
\newblock {Global Convergence of the EM Algorithm for Mixtures of Two Component
  Linear Regression}.
\newblock XX:1--57, 2019.

\bibitem{Leskovec2007}
Jure Leskovec, Andreas Krause, Carlos Guestrin, Christos Faloutsos, Jeanne
  VanBriesen, and Natalie Glance.
\newblock {Cost-effective Outbreak Detection in Networks}.
\newblock {\em Proceedings of the 13th ACM SIGKDD international conference on
  Knowledge discovery and data mining (KDD '07)}, page 420, 2007.

\bibitem{Meirom2014}
Eli~A. Meirom, Chris Milling, Constantine Caramanis, Shie Mannor, Ariel Orda,
  and Sanjay Shakkottai.
\newblock {Localized epidemic detection in networks with overwhelming noise}.
\newblock pages 1--27, 2014.

\bibitem{Milling2012}
Chris Milling, Constantine Caramanis, Shie Mannor, and Sanjay Shakkottai.
\newblock {Network Forensics : Random Infection vs Spreading Epidemic}.
\newblock In {\em Proceedings of the 12th ACM SIGMETRICS/PERFORMANCE joint
  international conference on Measurement and Modeling of Computer Systems
  (SIGMETRICS' 12)}, 2012.

\bibitem{Milling2015}
Chris Milling, Constantine Caramanis, Shie Mannor, and Sanjay Shakkottai.
\newblock {Local detection of infections in heterogeneous networks}.
\newblock {\em Proceedings - IEEE INFOCOM}, 26:1517--1525, 2015.

\bibitem{Myers2012}
Seth Myers, Chenguang Zhu, and Jure Leskovec.
\newblock {Information Diffusion and External Influence in Networks}.
\newblock In {\em Proceedings of the 18th ACM SIGKDD international conference
  on Knowledge discovery and data mining (KDD' 12)}, pages 33--41, 2012.

\bibitem{Netrapalli2012}
Praneeth Netrapalli and Sujay Sanghavi.
\newblock {Learning the Graph of Epidemic Cascades}.
\newblock In {\em Proceedings of the 12th ACM SIGMETRICS/PERFORMANCE joint
  international conference on Measurement and Modeling of Computer Systems
  (SIGMETRICS' 12)}, pages 211--222, 2012.

\bibitem{Newman2014a}
M.~E.~J. Newman.
\newblock {\em {Networks: An Introduction}}, volume~23.
\newblock 2014.

\bibitem{shah2010detecting}
Devavrat Shah and Tauhid Zaman.
\newblock {Detecting sources of computer viruses in networks: theory and
  experiment}.
\newblock In {\em ACM SIGMETRICS Performance Evaluation Review}, volume~38,
  pages 203--214. ACM, 2010.

\bibitem{Shah2010}
Devavrat Shah and Tauhid Zaman.
\newblock {Rumors in a Network : Who ' s the Culprit ?}
\newblock {\em IEEE Transactions on information theory}, 57(8):1--43, 2010.

\bibitem{shah2012rumor}
Devavrat Shah and Tauhid Zaman.
\newblock {Rumor centrality: a universal source detector}.
\newblock In {\em ACM SIGMETRICS Performance Evaluation Review}, volume~40,
  pages 199--210. ACM, 2012.

\bibitem{spencer2015impossibility}
Sam Spencer and R~Srikant.
\newblock {On the impossibility of localizing multiple rumor sources in a line
  graph}.
\newblock {\em ACM SIGMETRICS Performance Evaluation Review}, 43(2):66--68,
  2015.

\bibitem{wang2014rumor}
Zhaoxu Wang, Wenxiang Dong, Wenyi Zhang, and Chee~Wei Tan.
\newblock {Rumor source detection with multiple observations: Fundamental
  limits and algorithms}.
\newblock In {\em ACM SIGMETRICS Performance Evaluation Review}, volume~42,
  pages 1--13. ACM, 2014.

\bibitem{Wu2018}
Liang Wu and Huan Liu.
\newblock {Tracing Fake-News Footprints: Characterizing Social Media Messages
  by How They Propagate}.
\newblock In {\em (WSDM 2018) The 11th ACM International Conference on Web
  Search and Data Mining}, 2018.

\bibitem{Zarezade2017}
Ali Zarezade, Ali Khodadadi, Mehrdad Farajtabar, Hamid~R Rabiee, and Hongyuan
  Zha.
\newblock {Correlated Cascades : Compete or Cooperate}.
\newblock In {\em Proceedings of the Thirty-First AAAI Conference on Artificial
  Intelligence (AAAI-17)}, pages 238--244, 2017.

\bibitem{Zhao2015}
Qingyuan Zhao, Murat~A. Erdogdu, Hera~Y. He, Anand Rajaraman, and Jure
  Leskovec.
\newblock {SEISMIC: A Self-Exciting Point Process Model for Predicting Tweet
  Popularity}.
\newblock {\em Proceedings of the 21th ACM SIGKDD International Conference on
  Knowledge Discovery and Data Mining (KDD '15 )}, 2015.

\end{thebibliography}
\bibliographystyle{plain}
\newpage
\appendix
\section{Bidirectional tree} \label{app:tree}
We include here the full calculations for learning the weights of the bidirectional tree.

\begin{proposition}
If we know $(i,j)$ is an edge in the original tree, then the probability of infection along this edge is given by:
$$p_{ij} = \frac{f_{i<j}\cdot s_0 - f_{j<i}\cdot s_2}{g_{i,\bcancel{j}}\cdot(s_0^2 - s_2^2) + f_{i<j}\cdot s_0 - f_{j<i}\cdot s_2}.$$
\proof 
According to Lemma \ref{lem:exactProba}, we have:
\begin{align*}
    f_{i<j} &= \mathcal{P}_{\bcancel{j}}(\rightarrow i)\cdot p_{ij}\cdot s_{0} + \mathcal{P}_{\bcancel{i}}(\rightarrow j)\cdot p_{ji}\cdot s_{2} \\
    g_{i,\bcancel{j}} &= \mathcal{P}_{\bcancel{j}}(\rightarrow i)\cdot (1 - p_{ij}) \\
    f_{j<i} &= \mathcal{P}_{\bcancel{i}}(\rightarrow j)\cdot p_{ji}\cdot s_{0} + \mathcal{P}_{\bcancel{j}}(\rightarrow i)\cdot p_{ij}\cdot s_{2} \\
    g_{j,\bcancel{i}} &= \mathcal{P}_{\bcancel{i}}(\rightarrow j)\cdot (1 - p_{ji}).
\end{align*}
We have 4 second-order equations, with 4 unknowns: $p_{ij}$, $p_{ji}$, $\mathcal{P}_{\bcancel{j}}(\rightarrow i)$ and $\mathcal{P}_{\bcancel{i}}(\rightarrow j)$. We solve it:
\begin{align*}
    f_{i<j}  + f_{j<i} &= (\mathcal{P}_{\bcancel{j}}(\rightarrow i)\cdot p_{ij} + \mathcal{P}_{\bcancel{i}}(\rightarrow j)\cdot p_{ji})\cdot (s_{0} + s_{2}) \\
    f_{i<j} - f_{j<i} &= (\mathcal{P}_{\bcancel{j}}(\rightarrow i)\cdot p_{ij} - \mathcal{P}_{\bcancel{i}}(\rightarrow j)\cdot p_{ji})\cdot (s_{0} - s_{2}) \\
    \implies \mathcal{P}_{\bcancel{j}}(\rightarrow i)\cdot p_{ij} &= \frac{1}{2}\left(\frac{f_{i<j}  + f_{j<i}}{s_{0} + s_{2}} + \frac{f_{i<j} - f_{j<i}}{s_{0} - s_{2}} \right) \\
    &= \frac{f_{i<j}\cdot s_0 - f_{j<i}\cdot s_2}{s_0^2 - s_2^2} \\
    \mathcal{P}_{\bcancel{j}}(\rightarrow i) &= g_{i,\bcancel{j}} +  \mathcal{P}_{\bcancel{j}}(\rightarrow i)\cdot p_{ij} \\
    &= g_{i,\bcancel{j}} + \frac{f_{i<j}\cdot s_0 - f_{j<i}\cdot s_2}{s_0^2 - s_2^2} \\
   \implies p_{ij} &= \frac{\mathcal{P}_{\bcancel{j}}(\rightarrow i)\cdot p_{ij}}{\mathcal{P}_{\bcancel{j}}(\rightarrow i)} \\
   p_{ij} &= \frac{f_{i<j}\cdot s_0 - f_{j<i}\cdot s_2}{g_{i,\bcancel{j}}\cdot(s_0^2 - s_2^2) + f_{i<j}\cdot s_0 - f_{j<i}\cdot s_2}.
\end{align*}
\qed
\end{proposition}

\begin{lemma}
With $M = \frac{N^2}{\epsilon^2}\log\left(\frac{6}{\delta}\right)\frac{\left((s_0^2 - s_2^2 + s_0 + s_2)p_{max} + s_0+s_2\right)^2}{(s_0^2 - s_2^2)^2}$ samples, with probability at least $1-\delta$, we have:
$$|\hat{p}_{ij} - p_{ij}| \leq \epsilon.$$
\proof Using Hoeffding's inequality:
\begin{align*}
    \p(|\hat{f}_{i<j} - f_{i<j}| > \epsilon_1) &\leq 2e^{-2M\epsilon_1^2},\\
    \p(|\hat{f}_{j<i} - f_{j<i}| > \epsilon_1) &\leq 2e^{-2M\epsilon_1^2},\\
    \p(|\hat{g}_{i,\bcancel{j}} - g_{i,\bcancel{j}}| > \epsilon_1) &\leq 2e^{-2M\epsilon_1^2}.
\end{align*}
Choosing $M = \frac{1}{\epsilon_1^2}\log\left(\frac{6}{\delta}\right)$, we have that with probability at least $1-\delta$, all the following hold:
\begin{align*}
    |\hat{f}_{i<j} - f_{i<j}| &\leq \epsilon_1, \\
    |\hat{f}_{j<i} - f_{j<i}| &\leq \epsilon_1, \\
    |\hat{g}_{i,\bcancel{j}} - g_{i,\bcancel{j}}| &\leq \epsilon_1.
\end{align*}
Hence, with probability at least $1-\delta$, we have:
\begin{align*}
    \hat{p}_{ij} &= \frac{\hat{f}_{i<j}\cdot s_0 - \hat{f}_{j<i}\cdot s_2}{\hat{g}_{i,\bcancel{j}}\cdot(s_0^2 - s_2^2) + \hat{f}_{i<j}\cdot s_0 - \hat{f}_{j<i}\cdot s_2} \\
    &\leq \frac{(f_{i<j} + \epsilon_1)\cdot s_0 - (f_{j<i} - \epsilon_1)\cdot s_2}{(g_{i,\bcancel{j}} - \epsilon_1)\cdot(s_0^2 - s_2^2) + (f_{i<j} - \epsilon_1)\cdot s_0 - (f_{j<i} + \epsilon_1)\cdot s_2} \\
    &= \frac{f_{i<j}\cdot s_0 - f_{j<i}\cdot s_2 + \epsilon_1 (s_0 + s_2)}{g_{i,\bcancel{j}} \cdot(s_0^2 - s_2^2) + f_{i<j}\cdot s_0 - f_{j<i}\cdot s_2 -\epsilon_1(s_0^2 - s_2^2 + s_0 + s_2)} \\
    &= p_{ij}\frac{1}{1 - \frac{\epsilon_1(s_0^2 - s_2^2 + s_0 + s_2)}{g_{i,\bcancel{j}} \cdot(s_0^2 - s_2^2) + f_{i<j}\cdot s_0 - f_{j<i}\cdot s_2}} \\
    &\quad + \epsilon_1 \frac{s_0+s_2}{g_{i,\bcancel{j}} \cdot(s_0^2 - s_2^2) + f_{i<j}\cdot s_0 - f_{j<i}\cdot s_2} + o(\epsilon_1 ) \\
    &= p_{ij}\left(1 + \frac{\epsilon_1(s_0^2 - s_2^2 + s_0 + s_2)}{g_{i,\bcancel{j}} \cdot(s_0^2 - s_2^2) + f_{i<j}\cdot s_0 - f_{j<i}\cdot s_2}\right) \\
    &\quad + \epsilon_1 \frac{s_0+s_2}{g_{i,\bcancel{j}} \cdot(s_0^2 - s_2^2) + f_{i<j}\cdot s_0 - f_{j<i}\cdot s_2} + o(\epsilon_1 ) \\
    \hat{p}_{ij} - p_{ij} &\leq \epsilon_1\frac{(s_0^2 - s_2^2 + s_0 + s_2)p_{max}}{g_{i,\bcancel{j}} \cdot(s_0^2 - s_2^2) + f_{i<j}\cdot s_0 - f_{j<i}\cdot s_2} \\
    &\quad + \epsilon_1 \frac{s_0+s_2}{g_{i,\bcancel{j}} \cdot(s_0^2 - s_2^2) + f_{i<j}\cdot s_0 - f_{j<i}\cdot s_2} + o(\epsilon_1 ).
\end{align*}
Using the results from Lemma \ref{lem:exactProba}, we have:
\begin{align*}
    f_{i<j} &= \mathcal{P}_{\bcancel{j}}(\rightarrow i)\cdot p_{ij}\cdot s_{0} + \mathcal{P}_{\bcancel{i}}(\rightarrow j)\cdot p_{ji}\cdot s_{2}, \\
    g_{i,\bcancel{j}} &= \mathcal{P}_{\bcancel{j}}(\rightarrow i)\cdot (1 - p_{ij}).
\end{align*}
We use it to simplify the denominator:
\begin{align*}
    \text{denominator} &= g_{i,\bcancel{j}} \cdot(s_0^2 - s_2^2) + f_{i<j}\cdot s_0 - f_{j<i}\cdot s_2 \\
    &= \left(\mathcal{P}_{\bcancel{j}}(\rightarrow i)\cdot (1 - p_{ij}) \right)\cdot(s_0^2 - s_2^2) \\
    &\quad+ \left(\mathcal{P}_{\bcancel{j}}(\rightarrow i)\cdot p_{ij}\cdot s_{0} + \mathcal{P}_{\bcancel{i}}(\rightarrow j)\cdot p_{ji}\cdot s_{2}\right)\cdot s_{0}  \\
    &\quad- \left(\mathcal{P}_{\bcancel{i}}(\rightarrow j)\cdot p_{ji}\cdot s_{0} + \mathcal{P}_{\bcancel{j}}(\rightarrow i)\cdot p_{ij}\cdot s_{2}\right)s_2 \\
    &= \mathcal{P}_{\bcancel{j}}(\rightarrow i) \cdot(s_0^2 - s_2^2) \\
    &\geq \frac{s_0^2 - s_2^2}{N}.
\end{align*}
Plugging back above:
\begin{align*}
    \hat{p}_{ij} - p_{ij} &\leq \epsilon_1\frac{(s_0^2 - s_2^2 + s_0 + s_2)p_{max}}{g_{i,\bcancel{j}} \cdot(s_0^2 - s_2^2) + f_{i<j}\cdot s_0 - f_{j<i}\cdot s_2} \\
    &\quad + \epsilon_1 \frac{s_0+s_2}{g_{i,\bcancel{j}} \cdot(s_0^2 - s_2^2) + f_{i<j}\cdot s_0 - f_{j<i}\cdot s_2} + o(\epsilon_1 )  \\
    &\leq \epsilon_1 N\frac{(s_0^2 - s_2^2 + s_0 + s_2)p_{max}}{s_0^2 - s_2^2} + \epsilon_1 N\frac{s_0+s_2}{s_0^2 - s_2^2} + o(\epsilon_1 ).
\end{align*}
By symmetry, we obtain:
$$ |\hat{p}_{ij} - p_{ij}| \leq \epsilon_1 N \frac{(s_0^2 - s_2^2 + s_0 + s_2)p_{max} + s_0+s_2}{s_0^2 - s_2^2} + o(\epsilon_1 ).$$
By choosing $\epsilon_1 =  \frac{\epsilon}{N} \frac{s_0^2 - s_2^2}{(s_0^2 - s_2^2 + s_0 + s_2)p_{max} + s_0+s_2}$, we therefore have: \\
With $M = \frac{N^2}{\epsilon^2}\log\left(\frac{6}{\delta}\right)\frac{\left((s_0^2 - s_2^2 + s_0 + s_2)p_{max} + s_0+s_2\right)^2}{(s_0^2 - s_2^2)^2}$ samples, with probability at least $1-\delta$, we have $|\hat{p}_{ij} - p_{ij}| \leq \epsilon$.
\\\qed
\end{lemma}

\section{Bounded-degree graphs} 
\subsection{Solving the system}\label{app:bounded}

\begin{lemma}
Let $\Delta = \left(s_0^2 - s_2^2\right)^2 - 4(V_{ji}s_2 - V_{ij}s_0)(V_{ij}s_2-V_{ji}s_0)$. We have:
$$\Delta \geq (s_0^2 - s_2^2)^2\frac{(1 - p_{max})^2}{1 + p_{max}^2}.$$
\proof Finding a lower bound for $\Delta$ can be achieved through minimizing $\Delta$, or maximizing $(V_{ji}s_2 - V_{ij}s_0)(V_{ij}s_2-V_{ji}s_0)$. We want to solve:
\begin{equation*}
\begin{aligned}
& \underset{V_{ij}, V_{ji}}{\text{maximize}} & & (V_{ji}s_2 - V_{ij}s_0)(V_{ij}s_2-V_{ji}s_0) \\
& \text{subject to} & & V_{ij} = \frac{p_{ij}s_0 + p_{ji}s_2}{1 + p_{ij}p_{ji}}, \\
& & & V_{ji} = \frac{p_{ji}s_0 + p_{ij}s_2}{1 + p_{ij}p_{ji}}, \\
& & & p_{ij} \geq 0, \\
& & & p_{max} - p_{ij} \geq 0, \\
& & & p_{ji} \geq 0, \\
& & & p_{max} - p_{ji} \geq 0. \\
\end{aligned}
\end{equation*}
To do so, we introduce Lagrangian multipliers. By replacing $V_{ij}$ and $V_{ji}$ with their actual value, the optimization problem above only has affine constraints, so it satisfies the linearity constraint qualification for the Karush-Kuhn-Tucker conditions. In other words, all the partial derivatives of the Lagrangian are equal to 0 for an optimal point.
\begin{align*}
    \mathcal{L} &=\mathcal{L}(V_{ij}, V_{ji}, p_{ij}, p_{ji}, \lambda_1, \lambda_2, \mu_1, \mu_2, \mu_3, \mu_4) \\
    &= (V_{ji}s_2 - V_{ij}s_0)(V_{ij}s_2-V_{ji}s_0) \\
    &\quad- \lambda_1(v_{ij}(1 + p_{ij}p_{ji}) - p_{ij}s_0 + p_{ji}s_2) \\
    &\quad- \lambda_2(v_{ji}(1 + p_{ij}p_{ji}) - p_{ji}s_0 + p_{ij}s_2) \\
    &\quad- \mu_1p_{ij} - \mu_2 (p_{max} - p_{ij}) \\
    &\quad- \mu_3p_{ji} - \mu_4 (p_{max} - p_{ji}).
\end{align*}
We calculate the gradients of $\mathcal{L}$.
\begin{align*}
    \frac{\partial \mathcal{L}}{\partial V_{ij}} &= V_{ji}(s_0^2 + s_2^2) - 2V_{ij}s_0s_2 - \lambda_1(1 + p_{ij}p_{ji}), \\
    \frac{\partial \mathcal{L}}{\partial V_{ji}} &= V_{ij}(s_0^2 + s_2^2) - 2V_{ji}s_0s_2 - \lambda_2(1 + p_{ij}p_{ji}), \\
    \frac{\partial \mathcal{L}}{\partial p_{ij}} &= -\lambda_1(V_{ij}p_{ji} - s_0) - \lambda_2 (V_{ji}p_{ji} - s_2) -\mu_1 +\mu_2, \\
    \frac{\partial \mathcal{L}}{\partial p_{ji}} &= -\lambda_1(V_{ij}p_{ij} - s_2) - \lambda_2 (V_{ji}p_{ij} - s_0) -\mu_3 + \mu_4.
\end{align*}
From now on, we find the set $X^0$ of points for which all the partial derivatives are null. We know the solution of the maximization problem is the point of $X^0$ which maximizes the objective function.

Let us assume an interior point solution exists. For this point, all the gradients of $\mathcal{L}$ are equal to 0. Since it is an interior point, we also have $\mu_1 = \mu_2 = \mu_3 = \mu_4 = 0$ by complementary slackness. 
Solving this system, we obtain:
\begin{align*}
    \lambda_1  &= (s_0^2 - s_2^2)\frac{p_{ji}s_0 - p_{ij}s_2}{(1 + p_{ij}p_{ji})^2}, \\
     \lambda_2  &= (s_0^2 - s_2^2)\frac{p_{ij}s_0 - p_{ji}s_2}{(1 + p_{ij}p_{ji})^2}. 
\end{align*}
Plugging this in above, the condition $ \frac{\partial \mathcal{L}}{\partial p_{ij}} = 0$ becomes $p_{ij}p_{ji}(1 - p_{ji}) = 0$. However, this is impossible for an interior point, since $0 < p_{ij}, p_{ji} < p_{max} < 1$. Therefore, the extrema of $\Delta$ are attained when at least one constraint is active.

We notice that if the conditions $p_{ij} = 0$ or $p_{ji} = 0$ are active, then $(V_{ji}s_2 - V_{ij}s_0)(V_{ij}s_2-V_{ji}s_0) = 0$. Let us suppose (without loss of generality, by symmetry of the problem) that we have $p_{ij} = p_{max}$. The objective function is then increasing in $p_{ji}$. Therefore, $\Delta$ is minimized when $p_{ij} = p_{ji} = p_{max}$, which implies $v_{ij} = V_{ji} = \frac{p_{max}(s_0 + s_2)}{1 + p_{max}^2}$. In this case:
\begin{align*}
    \Delta &\geq (s_0^2 - s_2^2)^2 - 4V_{ij}^2(s_0 - s_2)^2\\
    &\geq (s_0^2 - s_2^2)^2 - 4\left(\frac{p_{max}(s_0 + s_2)}{1 + p_{max}^2}\right)^2(s_0 - s_2)^2 \\
    &\geq (s_0^2 - s_2^2)^2\left[ 1 - 4\frac{p_{max}}{1 + p_{max}^2}\right] \\
     &\geq (s_0^2 - s_2^2)^2\frac{(1 - p_{max})^2}{1 + p_{max}^2}.
\end{align*}
This expression is always positive, which is what we wanted.
\\\qed
\end{lemma}

\begin{theorem}
For any graph, for any noise distribution having finite values, we can learn the weights of all the edges of the graph. In particular, we can compute a quantity which converges to the true weight of each edge:
$$ \hat{p}_{ij} = \frac{2(\hat{V}_{ji}s_2 - \hat{V}_{ij}s_0)}{\left(s_0^2 - s_2^2\right) + \sqrt{\left(s_0^2 - s_2^2\right)^2 - 4(\hat{V}_{ji}s_2 - \hat{V}_{ij}s_0)(\hat{V}_{ij}s_2-\hat{V}_{ji}s_0)}}.$$
\proof 
\begin{align*}
    \hat{V}_{ij} &\to_{M\to\infty} V_{ij} \\
    &:= \frac{p_{ij}\cdot s_0 + p_{ji}\cdot s_2}{1+ p_{ij}\cdot p_{ji}} \\
    p_{ji} &= \frac{V_{ij} - p_{ij}\cdot s_0}{s_2 - V_{ij}\cdot p_{ij}}
\end{align*}
We can plug this in $V_{ji}$:
\begin{align*}
    V_{ji} &= \frac{p_{ji}\cdot s_0 + p_{ij}\cdot s_2}{1+ p_{ij}\cdot p_{ji}} \\
    V_{ji}\left[ 1 + p_{ij} \cdot \frac{V_{ij} - p_{ij}\cdot s_0}{s_2 - V_{ij}\cdot p_{ij}} \right] &= \frac{V_{ij} - p_{ij}\cdot s_0}{s_2 - V_{ij}\cdot p_{ij}}\cdot s_0 + p_{ij}\cdot s_2 
\end{align*}
After some shuffling around, we obtain the second-degree equation:
$$V_{ji}s_2 - V_{ij}s_0 + \left(s_0^2 - s_2^2\right) p_{ij} +\left(V_{ij}s_2-V_{ji}s_0 \right)p_{ij}^2 = 0  $$
We recall that by definition, $s_0 \geq s_2$. We also notice that if $p_{ij} = q_1$ and $p_{ji} = q_2$ is a pair of solutions of this system, then $p_{ij} = \frac{1}{q_2}$ and $p_{ji} = \frac{1}{q_1}$ forms the other pair of solution, which implies there is uniqueness of solutions in $[0,p_{max}]$. Since the real probabilities of infection satisfy this system, we also know the solution exists. Let $\Delta = \left(s_0^2 - s_2^2\right)^2 - 4(V_{ji}s_2 - V_{ij}s_0)(V_{ij}s_2-V_{ji}s_0)$. The only solution of this system in $[0,p_{max}]$ is:
\begin{align*}
    p_{ij} &=\frac{-\left(s_0^2 - s_2^2\right) + \sqrt{\Delta}}{2(V_{ij}s_2-V_{ji}s_0)} \\
    &= \frac{\left(s_0^2 - s_2^2\right)^2 - \left(s_0^2 - s_2^2\right)^2 - 4(V_{ji}s_2 - V_{ij}s_0)(V_{ij}s_2-V_{ji}s_0)}{2(V_{ji}s_0- V_{ij}s_2)\left(\left(s_0^2 - s_2^2\right) + \sqrt{\Delta}\right)} \\
    &= \frac{2(V_{ji}s_2 - V_{ij}s_0)}{\left(s_0^2 - s_2^2\right) + \sqrt{\Delta}}
\end{align*}
\qed
\end{theorem}

\subsection{Sample complexity}\label{sec:appBoundedComplexity}

\begin{proposition} 
As the number of cascades $M$ goes to infinity, the estimators below tend to the following limit:
\begin{align*}
    \hat{h}^2_{i,j} &\to_{M\to\infty} \frac{1}{N}(p_{ij} + p_{ji})\prod_{k \neq i,j} (1-p_{ik})(1-p_{jk}) \\
    \hat{f}^2_{i<j} &\to_{M\to\infty} \frac{1}{N}(p_{ij}\cdot s_0 + p_{ji}\cdot s_2)\prod_{k \neq i,j} (1-p_{ik})(1-p_{jk}) \\
    \hat{e}^1_i &\to_{M\to\infty} \frac{1}{N}(1-p_{ij})\prod_{k \neq i,j} (1-p_{ik})
\end{align*}
\proof Using the law of large numbers:
\begin{align*}
\hat{f}^2_{i<j} &\to_{M\to\infty} \E[\hat{f}^2_{i<j}] \\
&= \p(\text{$i$ source, infects $j$, no other infections, delay 0})\\
&\quad+\p(\text{$j$ source, infects $i$, no other infections, delay 2})\\
&= \frac{1}{N}p_{ij}\prod_{k \neq i,j} (1-p_{ik})\prod_{k \neq i,j}(1-p_{jk}) \cdot s_0\\
&\quad+ \frac{1}{N}p_{ji}\prod_{k \neq i,j} (1-p_{jk})\prod_{k \neq i,j}(1-p_{ik}) \cdot s_2 \\
&= \frac{1}{N}(p_{ij}\cdot s_0 + p_{ji}\cdot s_2)\prod_{k \neq i,j} (1-p_{ik})(1-p_{jk}).
\end{align*}
In the same vein, we have:
\begin{align*}
\hat{h}^2_{i,j} &\to_{M\to\infty} \E[\hat{h}^2_{i,j}] \\
&= \p(\text{$i$ source, infects $j$, no other infections})\\
&\quad+\p(\text{$j$ source, infects $i$, no other infections})\\
&= \frac{1}{N}(p_{ij} + p_{ji})\prod_{k \neq i,j} (1-p_{ik})(1-p_{jk}) \\
\hat{e}^1_i &\to_{M\to\infty} \E[\hat{e}^1_{i}] \\
&= \p(\text{$i$ source, no other infections})\\
&=\frac{1}{N}\prod_{k \neq i} (1-p_{ik}) \\
&=\frac{1}{N}(1-p_{ij})\prod_{k \neq i,j} (1-p_{ik}).
\end{align*}
\qed
\end{proposition}

\begin{proposition}
With probability $1- \frac{\delta}{N^2}$, with $M = $ samples, we can estimate $V_{ij}$ with precision $\epsilon_V$.
\proof As in Proposition \ref{lem:treeComplexity}, we use Hoeffding's inequality:
\begin{align*}
    \p(|\hat{f}^2_{i<j} - f^2_{i<j}| >\epsilon_1) &\leq 2e^{-2M\epsilon_1^2},\\
    \p(|\hat{h}^2_{i,j} - h^2_{i,j}| >\epsilon_1) &\leq 2e^{-2M\epsilon_1^2},\\
    \p(|\hat{e}^1_{i} - e^1_{i}| >\epsilon_1) &\leq 2e^{-2M\epsilon_1^2}, \\
    \p(|\hat{e}^1_{j} - e^1_{j}| >\epsilon_1) &\leq 2e^{-2M\epsilon_1^2}.
\end{align*}
We use this to bound above $V_{ij}$:
\begin{align*}
\hat{V}_{ij} &= \frac{\hat{f}^2_{i<j}}{\hat{h}^2_{i,j} + N\cdot \hat{e}^1_i \cdot \hat{e}^1_j} \\
&\leq \frac{f^2_{i<j} +\epsilon_1}{h^2_{i,j} +\epsilon_1 + N\cdot (e^1_i +\epsilon_1)\cdot (e^1_j +\epsilon_1)} \\
&= V_{ij} \frac{1 + \frac{\epsilon_1}{f^2_{i<j}}}{1 + \frac{\epsilon_1 + N\epsilon_1(e^1_i + e^1_j)}{h^2_{i,j} + Ne^1_i e^1_j} } \\
&= V_{ij} \left[ 1 + \frac{\epsilon_1}{f^2_{i<j}} +\epsilon_1\frac{1 + N(e^1_i + e^1_j)}{h^2_{i,j} + Ne^1_i e^1_j} + o(\epsilon_1) \right]
\end{align*}
We bound below the denominators:
\begin{align*}
    f^2_{i<j} 
    &\geq \frac{1}{N}(p_{ij}s_0 + p_{ji}s_2)(1-p_{max})^{2d} \\
    &\geq \frac{p_{min}s_2}{N} (1-p_{max})^{2d} \\
    h^2_{i,j} + Ne^1_{i}e^1_{j}  &\geq \frac{1}{N}(1 + p_{ij}p_{ji})(1-p_{max})^{2d}\\
    &\geq \frac{1}{N} (1-p_{max})^{2d}
\end{align*}
We bound above the numerator:
\begin{align*}
    N(e^1_{i} + e^1_{i}) &= N \left(\frac{1}{N}\prod_{k \neq i} (1 - p_{ik}) + \frac{1}{N}\prod_{k \neq j} (1 - p_{jk})\right) \\
    &\leq N \left(\frac{1}{N} + \frac{1}{N}\right) \\
    &\leq 2.
\end{align*}
Plugging in above:
\begin{align*}
\hat{V}_{ij} &= V_{ij} \left[ 1 + \frac{\epsilon_1}{f^2_{i<j}} +\epsilon_1\frac{1 + N(e^1_i + e^1_j)}{h^2_{i,j} + Ne^1_i e^1_j} + o(\epsilon_1) \right] \\
&\leq V_{ij} \left[ 1 + \frac{\epsilon_1}{\frac{p_{min}s_2}{N} (1-p_{max})^{2d}} + \frac{\epsilon_1(1 + 2)}{\frac{1}{N} (1-p_{max})^{2d}} + o(\epsilon_1) \right].
\end{align*}
Using $V_{ij}\leq 1$, and by symmetry:
\begin{align*}
|\hat{V}_{ij} - V_{ij}| &=  \epsilon_1\frac{N}{(1-p_{max})^{2d}}\left[\frac{1}{p_{min}s_2} + 3  \right] + o(\epsilon_1) \\
&= \leq \epsilon_1\frac{N}{(1-p_{max})^{2d}}\left[\frac{1 + 3p_{min}s_2}{p_{min}s_2}  \right] + o(\epsilon_1) \\
&\leq \epsilon_1\frac{4N}{p_{min}s_2(1-p_{max})^{2d}} + o(\epsilon_1).
\end{align*}
Therefore, by union bound, and by choosing  $\epsilon_1 \frac{4N}{p_{min}s_2(1-p_{max})^{2d}}= \epsilon_V$, and setting $2 e^{-2M\epsilon_1^2} = \frac{\delta}{3N^2}$, we obtain: \\
With $M =  \frac{1}{\epsilon_V^2} \frac{16N^2}{p_{min}^2s_2^2(1-p_{max})^{4d}}\frac{2\log(3N) - \log(\delta)}{2}$ samples, we can guarantee $|\hat{V}_{ij} - V_{ij}| \leq \epsilon_V $ with probability at least $1 - \frac{\delta}{N^2}$.
\qed
\end{proposition}

\begin{proposition} 
Assuming we can estimate $V_{ij}$ within precision $\epsilon_V$, then we can estimate $p_{ij}$ within precision $\epsilon = \frac{6\epsilon_V(1 + p_{max}^2)}{(s_0^2 - s_2^2)^2(1 - p_{max})^2}$.
\proof 
If we know $|\hat{V}_{ij} - V_{ij}| < \epsilon_V$ and $|\hat{V}_{ji} - V_{ji}| < \epsilon_V$:
\begin{align*}
    \hat{p}_{ij} &= \frac{2(\hat{V}_{ji}s_2 - \hat{V}_{ij}s_0)}{\left(s_0^2 - s_2^2\right) + \sqrt{\left(s_0^2 - s_2^2\right)^2 - 4(\hat{V}_{ji}s_2 - \hat{V}_{ij}s_0)(\hat{V}_{ij}s_2-\hat{V}_{ji}s_0)}} \\
    &\leq \frac{2(V_{ji}s_2 - V_{ij}s_0) + 2\epsilon_V(s_0 + s_2) }{\left(s_0^2 - s_2^2\right)^2 + \sqrt{\Delta  - 4\epsilon_V(s_0 + s_2)^2 (V_{ij}+V_{ji})}}.
\end{align*}
We recall $s_0 + s_2 \leq 1$, $V_{ij} \leq 1$, $p_{ij} \leq 1$, and $1 + p_{max}^2 \leq \Delta \leq 1$ (Lemma \ref{lem:delta}). Hence:
\begin{align*}
    \hat{p}_{ij} &\leq \frac{2(V_{ji}s_2 - V_{ij}s_0) + 2\epsilon_V }{\left(s_0^2 - s_2^2\right)^2 + \sqrt{\Delta}\sqrt{1 - 8\frac{\epsilon_V}{\Delta}}} \\
    &\leq p_{ij}\left( 1 + \frac{4\epsilon_V}{\sqrt{\Delta}\left( \left(s_0^2 - s_2^2\right)^2 + \sqrt{\Delta} \right)}\right) + \frac{2\epsilon_V }{\left(s_0^2 - s_2^2\right)^2 + \sqrt{\Delta}} + o(\epsilon_V) \\
    &\leq p_{ij} + \frac{6\epsilon_V}{\Delta}.
\end{align*}
By symmetry, and using the bound on $\Delta$ stated above, we conclude that if we know $V_{ij}$ and $V_{ji}$ up to precision $\epsilon_V$, we know $p_{ij}$ up to precision $\epsilon = \frac{6\epsilon_V(1 + p_{max}^2)}{(s_0^2 - s_2^2)^2(1 - p_{max})^2}$.
\\\qed
\end{proposition}

\begin{theorem}
In the limited-noise setting, with probability at least $1-\delta$, with $M =  \OO\left( \frac{e^{4p_{max}(d+1)}}{p_{min}^2s_2^2(s_0^2 - s_2^2)^4} \frac{N^2}{\epsilon^2} \log\left(\frac{N}{\delta}\right) \right)$ samples, we can learn the weights of any bounded-degree graph up to precision $epsilon$.
\proof With probability at least $1- \frac{\delta}{N^2}$, using \\$M =  \frac{1}{\epsilon_V^2} \frac{16N^2}{p_{min}^2s_2^2(1-p_{max})^{4d}}\frac{2\log(3N) - \log(\delta)}{2}$ samples, we can guarantee $|\hat{V}_{ij} - V_{ij}| \leq \epsilon_V $ with probability at least $1 - \frac{\delta}{N^2}$ samples, knowing $\frac{1}{\epsilon_V} = \frac{6(1 + p_{max}^2)}{\epsilon(s_0^2 - s_2^2)^2(1 - p_{max})^2}$. This gives us a sample complexity of:
\begin{align*}
    M &\geq \frac{18}{\epsilon^2(s_0^2 - s_2^2)^4(1 - p_{max})^{4(d+1)}} N^2 \left[\frac{4}{p_{min}s_2} \right]^2\log\left(\frac{9N^2}{\delta}\right) \\
    &\geq \frac{1152\cdot e^{4p_{max}(d+1)}}{p_{min}^2s_2^2(s_0^2 - s_2^2)^4} \frac{N^2}{\epsilon^2} \log\left(\frac{9N^2}{\delta}\right) \\
    &= \OO\left( \frac{e^{4p_{max}(d+1)}}{p_{min}^2s_2^2(s_0^2 - s_2^2)^4} \frac{N^2}{\epsilon^2} \log\left(\frac{N}{\delta}\right) \right).
\end{align*}
\qed
\end{theorem}
\end{document}